# Hybrid Wavelet and EMD/ICA Approach for Artifact Suppression in Pervasive EEG

Valentina Bono, Saptarshi Das, Wasifa Jamal, Koushik Maharatna

## Author's Emails:

vb2a12@ecs.soton.ac.uk (V. Bono*)

sd2a11@ecs.soton.ac.uk, s.das@soton.ac.uk (S. Das)

wj4g08@ecs.soton.ac.uk (W. Jamal)

km3@ecs.soton.ac.uk (K. Maharatna)

*Phone:* **+44-7708044466**

## Abstract

**Background**

Electroencephalogram (EEG) signals are often corrupted with unintended artifacts which need to be removed for extracting meaningful clinical information from them. Typically *a priori* knowledge of the nature of the artifacts is needed for such purpose. Artifact contamination of EEG is even more prominent for pervasive EEG systems where the subjects are free to move and thereby introducing a wide variety of motion-related artifacts. This makes hard to get *a priori* knowledge about their characteristics rendering conventional artifact removal techniques often ineffective.

**New Method**

In this paper, we explore the performance of two hybrid artifact removal algorithms: Wavelet packet transform followed by Independent Component Analysis (WPTICA) and Wavelet Packet Transform followed by Empirical Mode Decomposition (WPTEMD) in pervasive EEG recording scenario, assuming existence of no *a priori* knowledge about the artifacts and compare their performance with two existing artifact removal algorithms.

**Results**

Artifact cleaning performance has been measured using Root Mean Square Error (*RMSE*) and Artifact to Signal Ratio (*ASR*) – an index similar to traditional Signal to Noise Ratio (*SNR*), and also by observing normalized power distribution topography over the scalp.

**Comparison with Existing Method(s)**

Comparison has been made first using semi-simulated signals and then with real experimentally acquired EEG data with commercially available 19-channel pervasive EEG system Enobio corrupted by eight types of artifact.

**Conclusions**

Our explorations show that WPTEMD consistently gives best artifact cleaning performance not only in semi-simulated scenario but also in the case of real EEG data containing artifacts.

# 1. Introduction

Multi-channel EEG, due to its cost-effective and non-invasive nature, has been widely used in various clinical and commercial applications, starting from quantification of cognitive ability of a subject to aid diagnosis of neuro-degenerative diseases and, most recently, in Brain Computer Interface (BCI) application [1]. The most difficult part in dealing with EEG signals is the presence of artifacts that arise due to subject movements, physiological activity (e.g. respiration, cardiac and myogenic) and electrode contact problems. Unlike other physiological signals, e.g. Electrocardiogram (ECG), EEG does not have well-defined signal morphology and therefore it is often difficult to identify the artifacts uniquely from the actual EEG signal, since their frequency spectra often overlap; for instance, muscle activity is characterized by high amplitude, wide spectral distribution and variable topographical distribution [2]. This has triggered a whole body of research work to identify and suppress artifacts from the actual EEG signal. Typically, when the subject is at resting state, only a few artifacts can arise which could be well-characterized. Using this *a-priori* knowledge, several algorithms have been proposed for artifact removal from EEG following two main approaches:

(a) Measuring artifacts with supplementary sensors like Electro-oculogram (EOG) [10], and then applying linear filtering and regression to separate them from EEG;

(b) Blind Source Separation (BSS) techniques, like Independent Component Analysis (ICA) [3–5]. However, the fundamental point in all these approaches remains the same – existence of *a-priori* knowledge of the characteristics of the artifact based on which the algorithms are tuned to achieve maximal performance.

## *1.1. Motivation and Novelty*

In the recent years, wireless EEG systems [6] are becoming popular as they use dry contact electrodes which do not require conductive gel and skin preparation [7] resulting in reduced setup time for experiments and data acquisition. Since it allows studying the brain waves in unconstrained naturalistic settings [8], wireless EEG may consent assessment of cognitive functionalities of a subject [1], [9] during daily life. But such potential is hindered by the fact that, due to higher degrees of freedom of body movement, a wide variety of motion artifacts (e.g. head movement in yaw, pitch and roll, hand movement, talking and chewing, etc.) are introduced in recorded EEG. In practice, these movements strongly affect the recordings in such a way that the underlying EEG signal may not be recognizable. The biggest problem is that these artifacts are radically different from the traditional EEG literature (with high inter-trial variability) and being of random natures no *a-priori* knowledge exists about their characteristics, based on which they could be separated from EEG using the above mentioned approaches (see section 1.2 for more details). Moreover, the number of channels in wireless EEG is often less than the conventional EEG systems; therefore, the assumption that the clean EEG parameters should follow a normal distribution, used in formulating artifact separation algorithm like [10], could be violated because of low numbers of electrodes in pervasive EEG system.

In this paper, following the suggestion made by Uriguen [11], we explore the performance of two hybrid artifact removal techniques in the context of pervasive EEG with limited number of channels:

(a) Wavelet Packet Transform followed by Empirical Mode Decomposition (WPTEMD)

(b) Wavelet Packet Transform followed by ICA (WPTICA).

The main goal is to assess their performance in identifying and suppressing the artifacts corrupting the EEG signals without requiring any *a-priori* knowledge of the artifact characteristics, and consequently there is no possibility for tuning the thresholds of the algorithms as in [12]. Specifically, our exploration is targeted towards a pervasive unconstrained EEG scenario where the body movements are allowed. Our exploration started with semi-simulated dataset for benchmarking the performance of these algorithms against two state-of-the-art artifact separation methods based on thresholding criteria, namely wICA [12] and FASTER [10]. The relative performance was analyzed through the Root Mean Squared Error (*RMSE*) and a new quantitative index, called – Artifact to Signal Ratio (*ASR*) where the latter is analogous to the well-known Signal to Noise Ratio (*SNR*) used widely for quantifying signal quality. This exploration with semi-simulated data also helped in standardizing the above mentioned performance metric. Our results showed that these algorithms outperform [12] and [10] by 51.88% at recovering the EEG signal more accurately. Further exploration was carried out using commercially available pervasive EEG acquisition system Enobio [6] to capture EEG data corrupted with movement-related real-life artifacts. Even in this scenario, our results show that the WPTEMD is capable of reducing the artifact with minimal distortion in the original spectral characteristics of the signal.

The preliminary results of the current exploration have been reported in [13], with only four types of artifacts. Here we extend the analysis by:

(a) Exploring the validity of the parameters selected for the identification of the artifact used in these two hybrid techniques,

(b) Evaluating the algorithms with semi-simulated data to benchmark their performances against state-of-the-art algorithms,

(c) Analyzing quantitative performance of these algorithms on EEG signals corrupted with a larger variety of (eight movement-related) real-life artifacts.

### *1.2. Related Previous Work on Artifact Suppression*

Linear filtering and linear regression can separate an artifact from a corrupted signal if the artifact alone is measurable [14], e.g. eye-blinking artifact removal using EOG, muscle artifact removal using Electromyogram (EMG) and using Electrocardiogram (ECG) recoding for the artifact due to heart-beat [1], [9], [14]–[16]. In a pervasive EEG scenario, there is no known way to record the sources of artifact because of the higher degrees of freedom, so as to pose the artifact removal problem in linear filtering template [9]. Adaptive filtering techniques like Least-Mean-Square (LMS), Wiener filter in Finite/Infinite Impulse Response (FIR/IIR) form, Bayesian filtering (Kalman and particle filters) [9], [17] will need a statistical model of the artifact which is difficult to estimate due to high variability of the recordings over multiple trials and multiple subjects and the countless artifacts that could contaminate the recordings during natural body movement.

When the source of the artifact is unknown, *BSS* techniques like ICA [3]–[5], Principal Component Analysis (PCA) [18] and Canonical Correlation Analysis (CCA) [19]–[21] could be applied. In a pervasive EEG scenario, brain signals may be affected by a large amount of undetermined artifacts [22] which may be caused from distributed or multiple sources, rendering ICA and CCA less reliable for extracting the artifact components from EEG data, especially when dealing with low density EEG systems.

Contemporary researchers have introduced several *hybrid semi-automatic* techniques to tackle EEG contaminated by artifacts, e.g. wavelet enhanced ICA [12], [23]–[25], EMD-ICA [26], ensemble EMD (EEMD)-ICA [27], EEMD-CCA [28], deterministic (wavelet-EMD) and stochastic (ICA-CCA) approaches [29]. Akhtar *et al.* [23] propose an artifact reduction method based on Spatially Constrained ICA (SCICA) and wavelet denoising applied on EEG corrupted by eye-blinking artifact. They firstly apply SCICA [30], [31] to identify the artifactual independent components on which they later use the Stationary Wavelet Transform (SWT) to remove any leaked EEG activity and obtain only artifactual components. They, finally, reconstruct the EEG data by subtracting the identified artifact-only signals. The main drawback of this approach is that it is semi-automatic since it requires prior knowledge about the spatial topography characteristic of the sources generating the artifact. Castellanos *et al.* [12] provide a wavelet enhanced ICA (wICA) for removing ocular activity and heart beat from the corrupted EEG which was shown to perform better than ICA. Discrete Wavelet

Transform (DWT) is applied on the independent components extracted by the ICA; the wavelet coefficients exceeding a fixed threshold are set to zero and the signals are reconstructed. However, the selection of a threshold is an essential constituent of the algorithm, which needs tuning according to the peculiarities of the artifacts to remove. In this paper, we deal with real artifacts generated in natural environment that may be caused from several sources with unknown scalp topographies and, hence, do not allow us to follow Ahktar's [23] or Castellanos's [12] methods. As concluded by the authors [12], [23], these approaches are advantageous when applied on eye-blinking artifacts, characterized by well localized scalp topography but the feasibility of their application on other types of artifacts has not been investigated.

The wICA has been included in a Robust Artifact Removal (RAR) method, developed by Zima *et al.* [24], to remove short duration and high-amplitude artifacts from long-term neonatal 8-Channel EEG, mainly caused by movement activity. They segment the EEG for three times by following different rules and then wICA was applied on each of the three partitions. Instead of applying DWT on all the ICA components, they use a fixed threshold on the sparsity value of each component to identify the one containing the artifact which will be then decomposed by the wavelet and thresholded again as in wICA [12]. Later, they average the results from the three partitions and obtain the final artifact free EEG. Results showed that the method performs better than wICA. However, fine-tuning of the sparsity and wavelet thresholds are needed to apply the RAR method to other types of EEG recordings. Zima *et al.* [24] investigated the effect of different thresholds in wICA on the results and found that the default threshold is too aggressive and removes part of the actual EEG signal; this supports the need of an algorithm which does not depend on any threshold.

Safieddine *et al.* [29] compare two stochastic approaches (ICA and CCA) and two deterministic approaches (EMD and WT) in removing muscle artifacts in epileptic signals. Results showed that the performances depend on the level of contamination in the EEG signals. When data is highly corrupted by the myogenic artifacts, EMD outperforms the other methods, while ICA and WT give similar performances for less corrupted data. Because they have to deal with myogenic artifacts in epileptic signals, the authors make use of a threshold for: (a) the wavelet coefficients, (b) the Intrinsic Mode Functions (IMFs) for EMD and (c) the autocorrelation of the independent components to identify the artifactual component. In our case, since we are focusing on movements in a natural environment where both the nature of the EEG signal and the artifacts are different from the above, it is not feasible to find an ad-hoc fixed threshold suitable for all the possible artifacts contaminating the true EEG signals.

Sweeney *et al.* [28] combine the Ensemble Empirical Mode Decomposition with Canonical Correlation Analysis (EEMD-CCA) as a single channel technique to remove motion artifacts. They compare results obtained from two techniques employing wavelet [32] and EEMD-ICA [26] using two recordings of single channel EEG and functional near-infrared spectroscopy (fNRIS) data. To determine the artifactual component, they induce the motion artifact in one channel only, keeping the other channel as the 'ground truth' signal. In this way, they could identify the artifactual component as the one which, when removed, increased the correlation between the clean signal and the ground truth. Both EEMD-ICA and EEMD-CCA work as follows: firstly the single channel is decomposed by EEMD, several IMFs are then used as input for CCA and ICA to determine and remove the artifactual components. Finally, the single channel is reconstructed. This approach, as suggested by the authors, is not employable in a natural environment since it is not possible to record any ground truth signal but it was proved to be useful in comparing the algorithms. Another drawback of the method described above is even if the correlation coefficient is used within the channel during no motion epoch as a ground truth, dealing with multiple channels would mean calibrating for multiple thresholds for different types of artifacts.

Nolan *et al.* [10] develop a Fully Automated Statistical Thresholding for EEG artifact Rejection (FASTER) and test it on 128, 64 and 32-EEG channels ERP data collected during a visual oddball paradigm. The FASTER algorithm uses a subset of statistical thresholds on both EEG data and independent components obtained from ICA for the detection and rejection of channels and epochs affected by the artifacts. However, FASTER is based on the assumption that clean EEG parameters should be distributed normally and this could be violated in case of low numbers of data points. In fact, results showed that the algorithm can work effectively on all the datasets, including the 32-EEG channels, but the sensitivity of the algorithm decreases dramatically in case of low density EEG.

These problems motivated us to explore the possibility of designing an algorithm that performs artifact separation without *a-priori* knowledge of the artifacts (the real-life scenario in pervasive EEG applications) and corresponding threshold tuning. We followed the suggestion made in [32] to combine WPT, EMD and ICA in cascade to formulate WPTEMD and WPTICA algorithms and explore their comparative performance with the above mentioned goal in mind.

## 2. Material and Methods

In this section, we first give a brief description of the main constituents used in the two hybrid artifact suppression algorithms, followed by their combinations - WPTEMD and WPTICA. Later, we introduce the metrics used for comparing the different algorithms, along with the experimental protocol used for the generation of semi-simulated data and recording of real contaminated EEG and finally, we give a brief overview of the benchmarks wICA [12] and FASTER [10].

### 2.1. Wavelet Packet Transform (WPT)

WPT uses the DWT to decompose a signal by passing it through a series of high-pass and low-pass filter banks which allows the analysis of EEG signals in different scales and time resolutions. In WPT the decomposition is applied in both the detail $d[n]$ and approximate $a[n]$ coefficients, to get nodes at all the decomposed levels [33]. WPT has been previously applied as a feature extraction method to classify sleep stages on single EEG recordings [34] and in combination with EMD for the identification of continuous EEG during motor imagery tasks [35]. Although WPT is a single channel technique, we use it here from a multichannel perspective by considering the variance across multiple channels for each single WPT node. In fact, prior to any node removal, WPT is applied to all EEG electrodes in order to identify the most corrupted node (common to all channels) containing the artifact. For our algorithm Dmey Wavelet of Discrete Meyer family has been used. It has also been employed on EEG feature extraction in [36] in comparison with other family of mother wavelets. The signals are decomposed up to the seventh level, as used in Zima *et al.* [24]. In WPT, the frequency resolution increases and the temporal resolution decreases at each decomposition level. A trade-off between frequency and temporal resolution is required to successfully localize the artifact [23], hence in our experiments we empirically choose level seven decomposition which gives satisfactory artifact cleaning performance.

### 2.2. Independent Component Analysis (ICA)

ICA is a technique that uses the principle of statistical independence to find a representation in which the components of a signal are independent using the concept of non-Gaussian sources. Different ICA algorithms have been proposed in literature [37] which maximize the non-Gaussian measures like kurtosis or negentropy (differential entropy) although some of its variants are based on minimizing the mutual information. Infomax and FastICA [40] are the most common ICA algorithms: they showed similar performances when applied on ocular [38] and jaw clenching artifacts [39]. Here in the WPTICA hybrid algorithm we used FastICA [40], a highly computationally efficient method which implements a fixed-point iteration scheme for maximizing the negentropy as the chosen measure of non-Gaussianity. ICA has been traditionally used to identify the sources containing the artifacts, e.g. the eye-blinking artifact corrupting the frontal lobe electrodes [41].

### 2.3. Empirical Mode Decomposition (EMD)

EMD is a data driven decomposition technique for nonlinear and non-stationary signals. Given a signal $x(t)$, it can be decomposed as a linear combination of a finite number ($N$) of Intrinsic Mode Functions (IMFs) $h_i(t)$ and a residual $r(t)$, as in (1).

$$x(t) = \sum_{i=1}^{N} h_i(t) + r(t) \tag{1}$$

The IMFs should satisfy the following conditions [42]:

(a) The number of extrema and the number of zero-crossings must either be equal or differ at most by one.

(b) At any point the mean value of the envelope defined by the local maxima and local minima should be zero.

A smooth upper envelope $u(t)$ and a lower envelope $l(t)$ of the signal $x(t)$ are calculated from the extrema through interpolation. Then, the mean envelope $m(t)$ is calculated as $m(t) = [u(t) + l(t)]/2$ which is then subtracted from the original signal, i.e. $h(t) = x(t) - m(t)$. Once the $h(t)$ satisfies the conditions mentioned above, it will be subtracted from the signal and this sifting process will be repeated until the residual will be reached. A criterion is typically imposed to stop the sifting process, e.g. standard deviation $\sigma_s$ (0.2-0.3) between two consecutive shifts [42]. Imposing a low threshold may lead to over-iterations which results in over-decomposition. Another criterion was used in the WPTEMD algorithm, proposed by Rilling *et al.* [43]; it is based on two thresholds $\theta_1$ and $\theta_2$ to guarantee the presence of small global fluctuations (EEG) and large local excursions in the mean (artifact). Two parameters, mode amplitude $a(t)$ and evaluation function $\varepsilon(t)$, are estimated and compared with the predefined thresholds (2).

$$a(t) = \frac{u(t) - l(t)}{2}, \quad \varepsilon(t) = \left| \frac{m(t)}{a(t)} \right| \tag{2}$$

The sifting is repeated until the evaluation function $\varepsilon(t)$ will satisfy the following conditions with $\alpha \approx 0.05$, $\theta_1 \approx 0.05$ and $\theta_2 \approx 10\theta_1$.

$$\begin{cases} \varepsilon(t) < \theta_1 & \text{for } (1-\alpha) \text{ fraction of the signal length} \\ \varepsilon(t) < \theta_2 & \text{for the remaining fraction} \end{cases} \tag{3}$$

The decomposition achieved with EMD enables identifying the basic irregular components of the corrupted signal: the artifacts in the EEG can be identified as IMF and hence can be rejected to clean the signal.

### 2.4. Formulating Two Hybrid Artifact Suppression Algorithms: WPTEMD and WPTICA

In this paper, as a first step of suppressing the artifacts in pervasive EEG during natural body movement, we formulate a generalized framework for automatic removal of different artifacts, e.g. eye-blinking, head movement in yaw, pitch and roll, chewing, hand movement and talking using two hybrid approaches:

a) A combination of WPT and ICA (WPTICA)

b) A combination of WPT and EMD (WPTEMD)

Figure 1 shows the schematic representation of computation sequences of these two algorithms, the different parameters employed to quantify the effect of the artifact and the rules that are applied to automatically suppress the artifact in the corrupted channel.

Our aim is to explore the artifact suppression performance of these two algorithms particularly for a 19-channel EEG wireless system (Enobio). In both the methods, as shown in Figure 1, firstly the WPT is applied on the EEG data. The WPT decomposition up to level seven preserves the frequency granularity at the leaf nodes and also helps preserving most of the useful information of the original EEG. Therefore we considered up to $7^{th}$ decomposition-level for identifying the node containing the maximal effect of the artifacts. Since the EEG signal is modified due to the node removal of the WPT (equivalent to a band-stop filtering), rejecting a node of the $7^{th}$ level ensures that only a narrow-band of its frequency spectrum is filtered and most of the EEG information is retained. Although WPT is a single channel technique, we use it here as a multichannel technique by computing the variation in energy across WPT tree for all the 19 EEG channels, prior to any node removal. The energy of the wavelet coefficients for each node of the 19 trees obtained is calculated next. To identify the presence of the artifact, we aimed at finding a node common to all channels which maximally captures the effect of the artifact. The criterion used is the maximal standard deviation of the wavelet energy $\sigma_{EGY}$ across all channels, as described in (4).

$$\sigma_{EGY,node_i} = \frac{1}{C-1} \sum_{c=1}^{C} \left( \text{EGY}_{c,\,node_i} - \overline{\text{EGY}}_{\text{across channels, node}_i} \right)^2,$$

$$\text{EGY}_{c,\,node_i} = \sum_{n=1}^{N} \left( node_i[n] \right)^2,$$

$$i \in \{1,2,\cdots,I\} = node, \ c \in \{1,2,\cdots,C\} = \#\,of\ channels,$$

$$I = 2^7 = 128, C = 19, N = \#\,of\ sample\ at\ the\ node$$

(4)

The coefficients of that particular node (characterized by the maximum $\sigma_{EGY}$) are then rejected while reconstructing the artifact suppressed EEG signal. The obtained WPT-cleaned signal is fed to the ICA (WPTICA) or EMD (WPTEMD) for further suppression of the artifact, as shown in Figure 1.

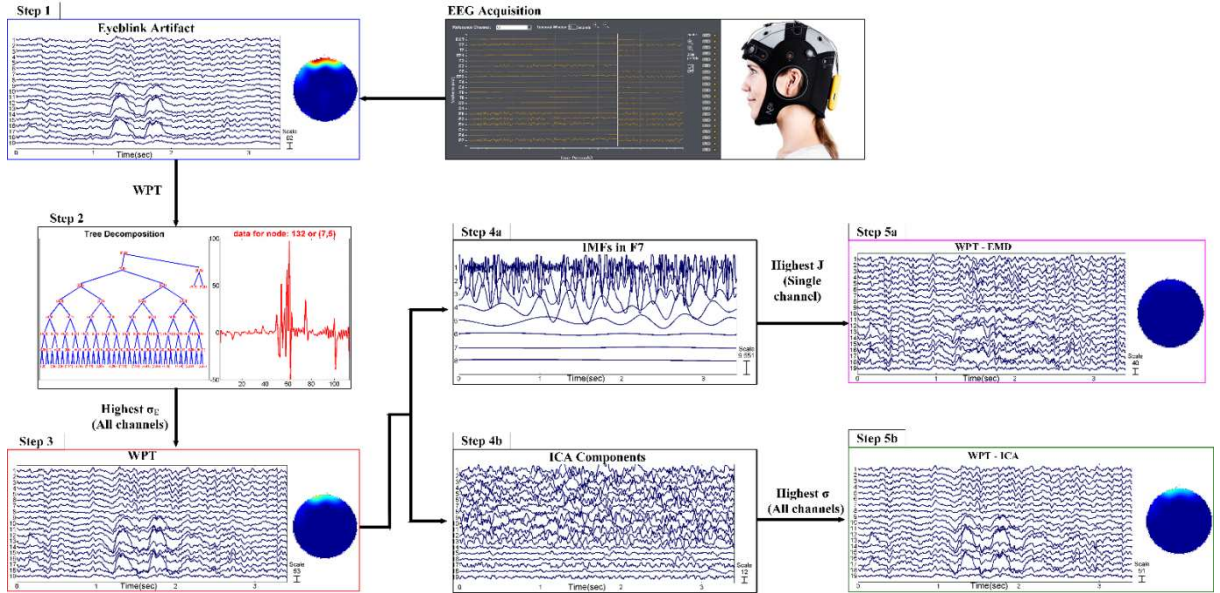

Figure 1: Schematic diagram of computation sequence of the two hybrid algorithms WPTEMD and WPTICA. The steps 2 and 3 are common to both the algorithms.

The *WPTICA* algorithm employs the FastICA routine [40] on the 19-channels EEG with the objective of separating the common component of the artifact traced in all the electrodes. The temporal standard deviation $\sigma_c$ of each independent component within the artifact region is used as a criterion to identify the component containing the artifact. This is characterized by the highest $\sigma_c$ and it is considered as the most influential part of the artifact, affecting some/all of the channels and hence rejected during the signal reconstruction.

When using *WPTEMD* technique, EMD is applied as a second step on the WPT-cleaned signal instead of ICA. The EMD, being a nonlinear signal decomposition technique, tries to capture the irregular oscillations with inconsistently large amplitude. These are the characteristics of all types of artifacts investigated here. The IMF with the highest parameter *J* is considered the one which fulfills the above criteria. The *J* criterion in (5) has been formulated as a weighted sum of normalized entropy *H* and standard deviation $\sigma$ with respect to their resting state values as shown in Figure 1. These resting state parameters are extracted from resting EEG, i.e. while the subject has the eye-closed and is not performing any task or movement.

$$J = w\left(H/H_{resting}\right) + (1-w)\left(\sigma/\sigma_{resting}\right) \tag{5}$$

### 2.5. Parameter Selection for the Index J of WPTEMD

The IMF with the highest *J* parameter is considered as the most accountable part in the artifact region, hence it is rejected during the reconstruction of the cleaned signal. The entropy *H* and the standard deviation $\sigma$ of each IMF in both the corrupted and the resting state EEG are used to evaluate the hybrid index *J*, described in (5), for each channel to capture large inconsistent low frequency oscillations as IMFs and potential components of the artifact.

The information entropy *H* [44], described in (6), captures the large amount of randomness introduced by the artifact in the IMFs and helps rejecting the corrupted component in WPTEMD hybrid algorithm.

$$H\left(IMF_i\right) = -\sum_{l=1}^{L}\left(IMF_i^2[l]\log\left(IMF_i^2[l]\right)\right) \tag{6}$$

$l = 1, \cdots, L, L = \#$ of samples in the IMF

Shannon entropy has already been used to detect unusual activity patterns in EEG data on ICA components [45], since it is a measure of randomness of a signal. The temporal standard deviation $\sigma$ of the IMFs captures the effect of large inconsistent fluctuations due to the artifact compared to that of the EEG (characterized by low amplitude high frequency oscillations). The combined use of standard deviation $\sigma$ and entropy *H* enables us to take into account different types of artifacts characterized by both higher randomness, like muscle activity [46] and large spikes, like eye-blink or motion artifacts. A parameter *w* was used to weight the normalized *H* and $\sigma$. Its value has been empirically found equal to 0.5. We used eye-blink artifact for different subjects and trials to identify the optimal value of *w*. The eye-blink was chosen for the selection of *w*, since its scalp topography is well known and *J* parameter should be higher in the frontal electrodes (the most corrupted). Figure 2 shows an example of *J* values obtained by varying *w* in a single subject-single trial, while Figure 3 shows that *J* values (with *w* = 0.5) are indeed higher in the frontal electrodes for the eye-blink artifact across all subjects and all trials. The *J* parameter was calculated by varying the weight *w* which balances the relative weight between the normalized *H* and the normalized $\sigma$ to find out the more significant contribution in the identification of the corrupted IMF.

Figure 2 shows an example (single subject single trial of artifact) of the *J* parameter for eye-blinking artifact while varying the weight *w* to select its optimum value. Highest values of the *J* index are found in IMF3 and IMF4, regardless of the value of *w*, and indicate the presence of the artifact. From this example, it is visible that the IMF3 which is identified as artifact (with maximum *J*), does not change if a different value of *w* or any other electrode is chosen. Since both the normalized *H* and $\sigma$ identify the same IMF as containing the artifact, the weight parameter *w* provides equal scaling of the two parts of *J* and hence chosen as 0.5. Similar results are reported for other types of artifact in the supplementary material.

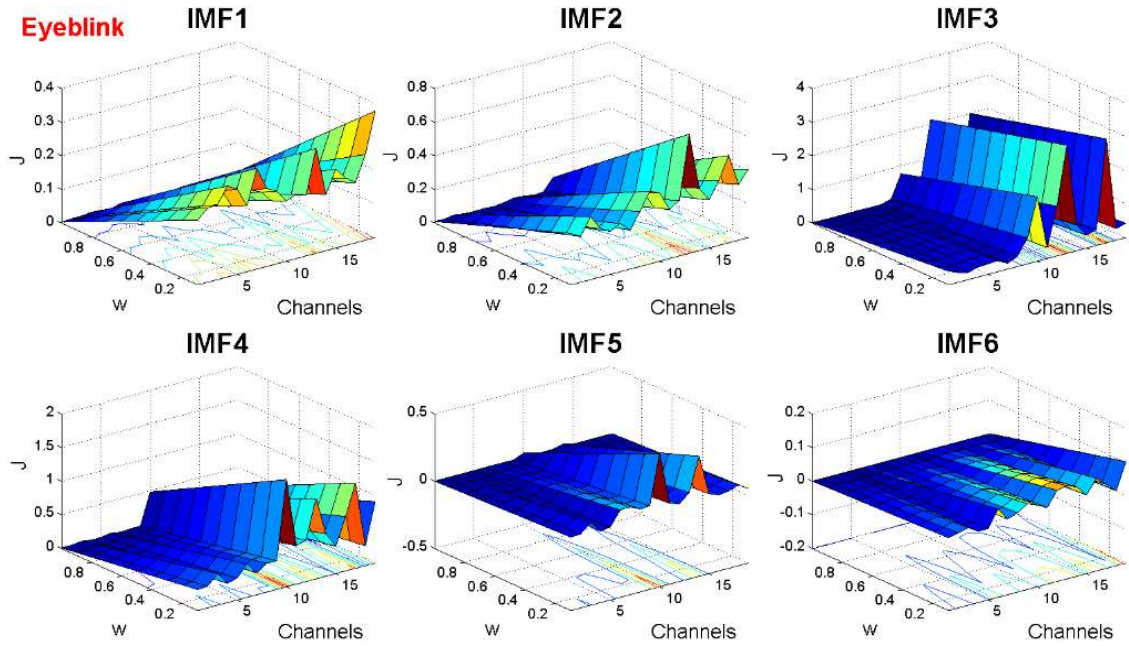

Figure 2: Surface plot of the values of the $J$ parameter across channels and IMFs for a single trial eye-blink artifact for different values of w ranging between 0 and 1.

The value of the $J$ index has been evaluated across different subjects and trials for eye-blinking experiments. Figure 3 shows that, as expected, the electrodes in the frontal region show the highest median value and the highest variability or Interquartile Range (IQR), indicating the presence of the artifact. The appropriate choice of the $J$ index is confirmed by these results, since the typical signature of the effect of the eye-blink is related to the presence of the artifact in the frontal electrodes. As shown for the single subject case of eye-blink artifact in Figure 2, the group analysis in Figure 3 for the same artifact also identifies the IMF3 as the most affected one (median>2), especially in the two channels Fp1 and Fp2. Additional analysis of the variability of $J$ for other types of artifact are provided in the supplementary material. When considering all the trials for each type of artifact, mostly the IMF4 is rejected as shown in the Table I in the supplementary material.

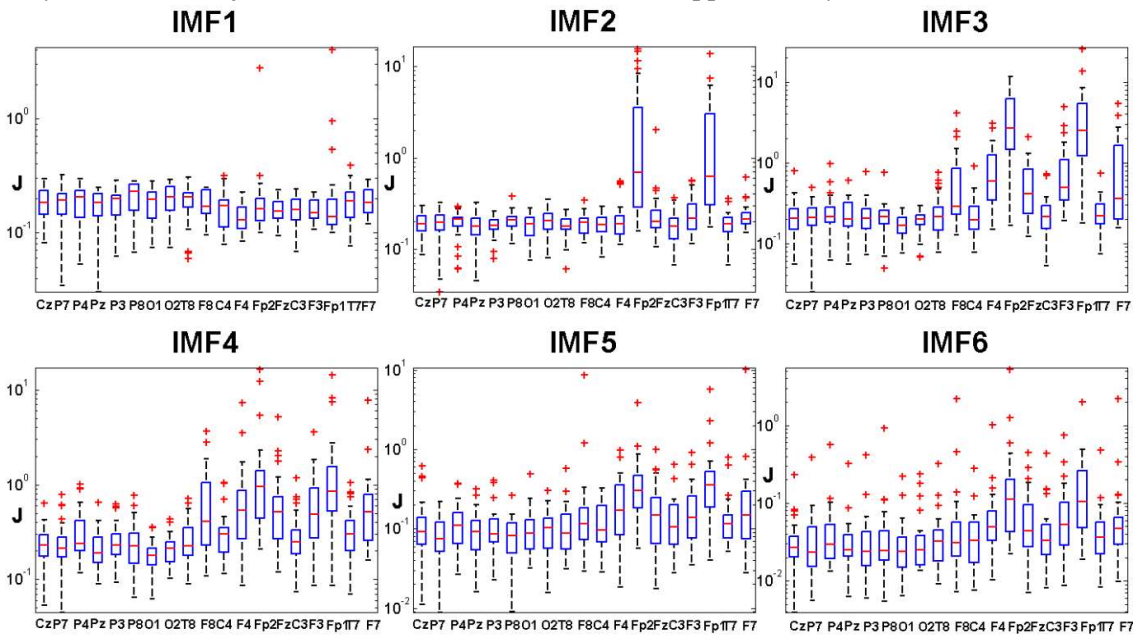

Figure 3: Distribution of the $J$ parameter for each IMF and across all trials and subjects for the eye-blink artifact. Consistently with the scalp topography of the eye-blink, $J$ parameter shows higher median values in the frontal electrodes.

### *2.6. Performance Evaluation*

Results from both the algorithms have been evaluated and compared by:

(a) Visual inspection of time domain morphology of the signal

(b) Frequency spectrum

(c) Spatial scalp topography of the artifact processed EEG data, as suggested in [11].

The visual inspection in time domain helps us to investigate the quality of the recovered brain signals after the artifact removal, while looking at the power spectrum aids to inspect the distortions of the EEG power spectrum. In fact, as discussed in [12], it is worth investigating how the spectrum of the underlying EEG activity is altered by the artifact separation algorithms through a qualitative analysis of the distortions in the power spectrum and the scalp topography.

Two criteria have been used here in quantitative performance evaluation for artifact removal by the two hybrid algorithms:

(a) Root Mean Squared Error (*RMSE*) criterion

(b) Artifact to Signal Ratio (*ASR*) criterion.

The *RMSE* quantifies the similarity between the original data $z_1$, without the artifact, and the artifact processed signal $z_2$ for each channel. In case of semi-simulated data, the original data $z_1$ without artifact are considered as our ground truth; hence the *RMSE* can be calculated as in (7).

$$RMSE = \sqrt{\frac{1}{K}\sum_{k=1}^{K}\left[z_1(k) - z_2(k)\right]^2} \tag{7}$$

Since in case of experimentally acquired data the ground truth is unknown, to quantify the extent of artifact cleaning for a real EEG dataset, we here introduce a metric termed as *ASR*. It is a new formulation of the SNR-like criterion [13] to better understand the cleaning performances of the two artifact removal methods in each EEG frequency band (i.e. $\delta$, $\theta$, $\alpha$, $\beta$ and $\gamma$). The mathematical form of *ASR* is described in (8) and is applied to each electrode for both the hybrid algorithms.

$$ASR = P_{Artifact}/P_{Resting} = \left(P_{Corrupted} - P_{Clean}\right)/P_{Resting} \tag{8}$$

The resting state is acquired while the subject has the eyes closed and the artifacts are absent. The corrupted signal is acquired while the subject performs a specific type of motion task. While calculating the *ASR*, the power of the artifact is obtained from the difference (in power) between the corrupted and the clean signals. For evaluating the power of the resting state EEG, only one trial was considered. Higher values of *ASR* indicate better performances: since the $P_{Resting}$ and $P_{Corrupted}$ are constant, lower value of $P_{Clean}$ will increase the numerator, which results in an increased *ASR*. Low values of *ASR* indicate that the power of the clean signal is comparable to the power of the corrupted one, suggesting that the algorithm did not modify the power spectrum of the corrupted signal in that particular frequency band. *ASR* will have negative values when the power of the 'cleaned' signal is increased after the algorithm has been applied.

In case of semi-simulated data, to understand whether the *ASR* can quantify the amount of artifact on the corrupted EEG, we evaluate an *ASR* ground truth ($ASR_{GT}$) denoted in (9) and subtracted it from the *ASR* formulated in (8), to obtain $\Delta ASR = ASR - ASR_{GT}$:

$$ASR_{GT} = P_{Artifact}/P_{EEG} \tag{9}$$

where $P_{ARTIFACT}$ and $P_{EEG}$ indicate respectively the power of the artifact and the EEG described in (10).

Since $ASR_{GT}$ signifies the ground truth of the artifact to signal power ratio, $\Delta ASR$ close to zero indicates the method with the least distortion of the power spectrum of the reconstructed EEG signal.

As mentioned above, semi-simulated data are used here to obtain preliminary results on ground truth signals as well as to standardize *ASR* measure and to compare and analyze the performance of these algorithms. Once established through semi-simulated data, *ASR* is then used to compare the WPTEMD and WPTICA algorithms performance for eight types of real artifacts, resembling a natural movement while the EEG is being acquired.

### 2.7. Generation of Semi-simulated Data

The two algorithms have been applied on data containing semi-simulated artifacts. 32 trials of 6 sec artifact free-EEG epochs were extracted from the resting data of different subjects by careful visual inspection and acquired with 19-channel Enobio wireless EEG system [6]. Synthetic eye-blink artifacts have been generated by Matlab functions developed in [47] and multiplied by a typical eye-blink scalp topography (high gain on the frontal electrodes and near zero for the rest) extracted from a different subject to obtain 19-channel EEG data contaminated by eye-blink, as in [48]. We generated 32 trials of each artifact with a sampling rate of 500 Hz and a length of 6 sec. Different levels of artifact contamination were produced by adding the artifact component $Artifact(t)$ to the EEG signals $EEG(t)$ [49], as follows:

$$x(t) = EEG(t) + \lambda \times Artifact(t) \tag{10}$$

where $\lambda$ represents the contribution of the artifact contamination and it varies between 1 and 15 to estimate the performance of the algorithms for different extent of contamination. Figure 4 shows an example of the original resting EEG data (*EEG only*), corrupted (*Corrupted*) with simulated eye-blink artifact ($\lambda = 10$) and reconstructed with all the algorithms under investigation.

However, since the insertion of simulated artifacts may not correspond to the real contamination of the EEG data and due to the unavailability of existing database containing wide variety of real EEG artifact examples, the acquisition of real data was needed, as described in the following subsection.

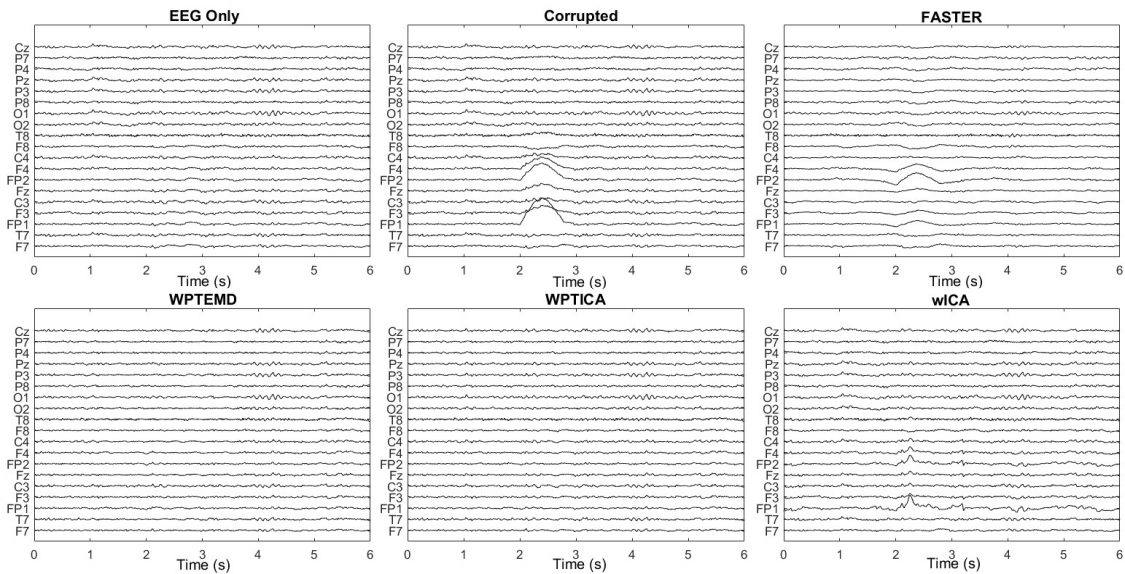

Figure 4: Example of semi-simulated EEG data for a single trial with artifact contamination level of $\lambda = 10$. The 'EEG Only' is the EEG recording prior to artifact contamination, Corrupted is the EEG recording contaminated by the simulated artifact, the rest of the EEG data are obtained by processing the corrupted with the corresponding techniques.

### 2.8. Experimental Protocol for Data Acquisition Using Pervasive EEG Cap

Real-life data was acquired using the 19-channel Enobio wireless EEG system [6] from 10 subjects, three females and seven males with mean age of 28.8 years (standard deviation 3.05). The participants were seated approximately 80 cm from a computer monitor with backrest and armrest and were asked to perform movements with the purpose of creating eight different types of artifacts, in order to obtain noisy signals that could be later investigated. Each type of movement was repeated over three trials for each subject. The Enobio system allows marking of the desired portion of EEG, thus labelling of the type of artifacts during the acquisition. Subjects were asked to perform the following tasks for

three trials:

1) *Resting state*: performed in the first and the last minute of the experiment. It is defined when the subject's eyes are closed and no task is being performed [50]
2) *Eyes open*: keeping the eyes open until a specific task was verbally instructed
3) *Eye-blinking*: it is a natural blink without eyes and head movement
4) Head movement from left to right (*yaw*)
5) Head movement from up to down (*pitch*)
6) Head shaking (*roll*)
7) *Hand movement* (left only and right only): one hand at a time was moved along the border of a tablet screen
8) *Chewing*: it involves most of the facial muscles; it was performed for five seconds per each trial
9) *Hello*: the volunteers were asked to say the word 'hello' to acquire artifact generated while speaking.

### 2.9. Comparison with Benchmarks Algorithms

To evaluate the performance of these methods in comparison to the state-of-the-art, we consider two competitive benchmarks consisting of wICA [12] and FASTER [10]. The *wICA* combines the use of the DWT and ICA to suppress the EEG data contaminated by the artifacts. Each independent component is decomposed by the DWT to identify the wavelet coefficients containing the artifact: those coefficients are identified by a fixed threshold and set to zero. The independent components are then reconstructed to obtain artifact-free EEG data.

*FASTER* is an automated artifact removal method based on thresholding criteria to detect and suppress bad channels, epochs and subject's data heavily contaminated by artifacts. It is based on different steps, each of which estimates and thresholds various statistical parameters, e.g. variance, mean correlation, Hurst exponent, amplitude range, channel deviation, spatial kurtosis, correlation with EOG reference electrode and median deviation with the aim of:

(a) Identify the contaminated channels to reject and substitute them with the interpolation of neighbor electrodes
(b) Identify and remove the contaminated epochs
(c) Detect and subtract the contaminated independent components extracted by ICA
(d) Detect and remove bad channels within the epochs
(e) Remove subject's data heavily contaminated by artifacts.

## 3. Results and Discussion

In this section, we first compare the performance of the proposed hybrid algorithms with the benchmark algorithms in terms of *RMSE* and *ASR* with the semi-simulated data. Experimentally acquired data corrupted with real-life artifacts is then used to explore the performances of the algorithms under consideration in terms of *ASR*, scalp topography and through a qualitative analysis in time domain morphology and frequency spectrum.

### 3.1. Analysis of the Semi-Simulated Artifact

Here we present simulation results to quantitatively analyzing efficiency of the WPTEMD and WPTICA, in comparison with the benchmarks FASTER[1] [10] and wICA[2] [12].

---

[1] The implementation of FASTER algorithm was freely available at http://www.mee.tcd.ie/neuraleng/Research/Faster
[2] The implementation of wICA algorithm was freely available at http://www.mat.ucm.es/~vmakarov/downloads.php

Figure 4 shows the results obtained with the four algorithms in case of semi-simulated artifact ($\lambda$ = 10): the frontal electrodes are highly corrupted and the artifact is still present in the EEG reconstructed by wICA and FASTER, while the WPTEMD and WPTICA suppress the contamination.

Figure 5 compares the performances of the four algorithms (2 proposed and 2 benchmarks) in terms of *RMSE* for eye-blinking artifact in: (a) two of the most affected channels (i.e. Fp1 and Fp2) and (b) in one of the channels less contaminated by the artifact (i.e. O$_2$) across the 32 trials and for different standard deviation of the artifact (given by $\lambda$ in (10)). When applied on heavily corrupted channels (e.g. Fp1 and Fp2), as shown in Figure 5, both the algorithms under exploration outperform the wICA in terms of *RMSE* when $\lambda > 5$ and their advantages become more prominent as $\lambda$ increases (i.e. as the extent of contamination increases) from 6 to 15. We performed a left tailed paired t-test to compare the performance of the algorithms. In the case of channel Fp1, the respective performance improvement of WPTEMD and WPTICA against wICA is 41.40% (p = 0.00059) and 41.50% (p = 0.00048) for $\lambda$ = 10 which increases to 51.88% (p = 0.00021) and 51.75% (p = 0.00018) for $\lambda$ = 15. On the other hand, in Fp2 these performance enhancements reach 49.53% (p = 0.0002) and 46.03% (p = 0.00053) for WPTEMD and WPTICA respectively when $\lambda$ = 15. It is worth mentioning that for highly contaminated channels, $\lambda$ = 10 corresponds approximately to an amplitude range of 400 $\mu V$, which is the case of heavily contaminated data due to subject's body movement in pervasive scenario. Figure 5 also shows that for Fp1 the performance of FASTER, WPTEMD and WPTICA are all equivalent over the entire range of $\lambda$. However, for the channel Fp2 the performances of WPTEMD and WPTICA are significantly better than FASTER - 70.40% (p = $1.09 \times 10^{-28}$) and 68.35% (p = $1.20212 \times 10^{-26}$) respectively for $\lambda$ = 15. A similar trend can also be observed for the channel O$_2$ where the artifact contamination is very low (see the time-domain signal morphology in Figure 4). But on the other hand, in terms of *RMSE*, for such low-corrupted channel the best performance has been achieved by the wICA as it is also evident from Figure 5.

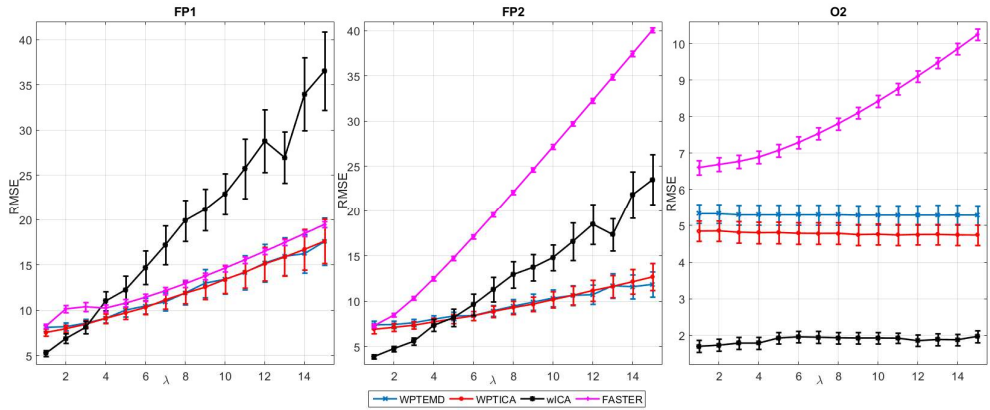

Figure 5: RMSE for the semi-simulated EEG data in channels Fp1, Fp2 and O2 across all 32 trials for different strengths ($\lambda$) of the artifact. The error bars represent one standard error of the mean ($\mu \pm \sigma$) of RMSE.

Figure 6 shows the $\Delta ASR$ (i.e. difference between the $ASR$ and the $ASR_{GT}$) in each frequency band for WPTEMD, WPTICA, FASTER and wICA reconstructed signals. In the performance comparison in terms of $\Delta ASR$, a $\Delta ASR$ closer to zero indicates better performance as explained in Section 2.6. It is apparent from Figure 6 that $\Delta ASR$ for WPTEMD remains close to zero and +ve in all the bands and for any value of $\lambda$, whereas $\Delta ASR$ for the other algorithms varies as the value of $\lambda$ increases, in some cases becoming –ve indicating a change in the power spectrum of the reconstructed signal. In $\delta$ band, which is the most contaminated band by the artifact, for high $\lambda$ ($\lambda$>5) both the WPTEMD and WPTICA can be considered the best methods in terms of $ASR$, since the $\Delta ASR$ is closer to zero, being the $\Delta ASR$ equals to 0.01 for WPTEMD, -0.1 for WPTICA, -1.82 for wICA and -3.74 for FASTER. In case of $\theta$ band for high $\lambda$ ($\lambda$>9), WPTEMD outperforms all the other methods in terms of $\Delta ASR$. In the $\alpha$ band, wICA and WPTICA perform similarly and show the $\Delta ASR$ values closest to zero, even though the negative value of $\Delta ASR$ indicates that these two methods modify the power spectrum of the original signal in this band. In case of $\beta$ band, the wICA performs better than the rest of the methods; while in $\gamma$ band both wICA and WPTEMD show similar performances. It is worth noting that WPTEMD is designed for heavily contaminated EEG data, in which cases (frontal channels and high $\lambda$) has shown the best performances in terms of $RMSE$ and $ASR$ for semi-simulated eye-blink.

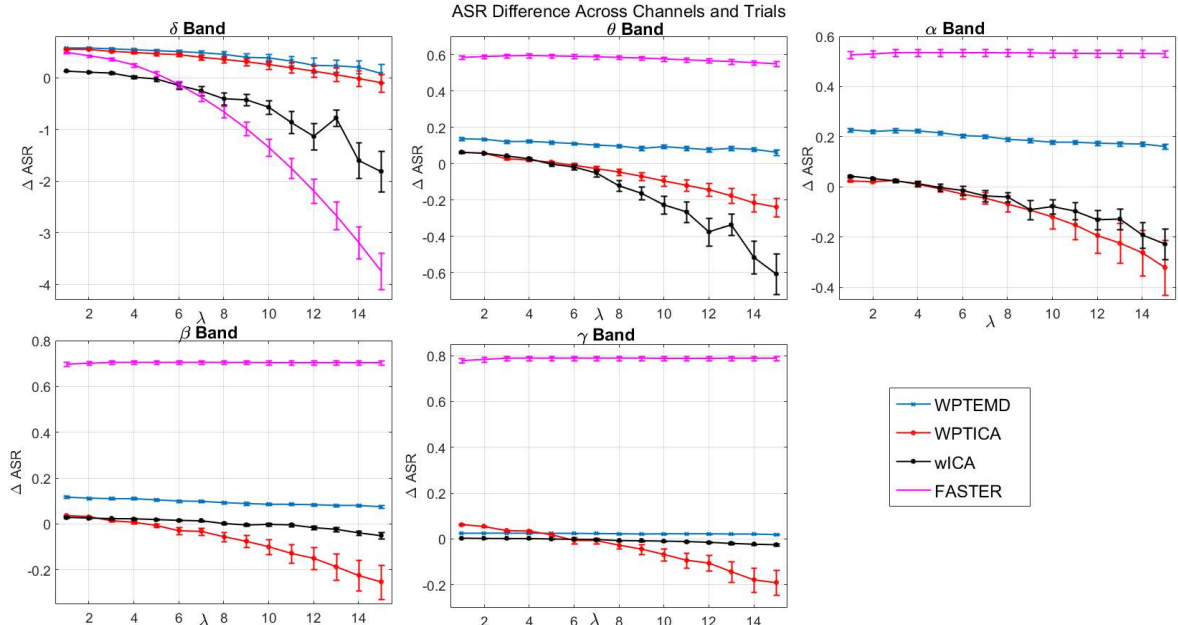

Figure 6: $\Delta$ASR in each frequency band of the semi-simulated artifact, across different trials and channels for different strengths ($\lambda$) of the artifact. The error bars represent one standard error of the mean ($\mu \pm \sigma$) of $\Delta$ASR.

Figure 7 shows the performance of all the four algorithms on the semi-simulated artifact, for a single trial in time domain as well as in the respective frequency bands. Since the eye-blink affects mostly the frontal electrodes, the EEG data in channel Fp1 is shown as an example for $\lambda = 15$. The time domain graph shows that all techniques reduce the large oscillation caused by the artifact, but the visual appearance of the WPTEMD-reconstructed signal is more similar to the original artifact-free EEG. In addition, the power spectra in different bands show that, especially above 6 Hz the WPTEMD and wICA give the closest spectrum to the original one, while WPTICA and FASTER overestimate and underestimate the spectra of the reconstructed signal respectively.

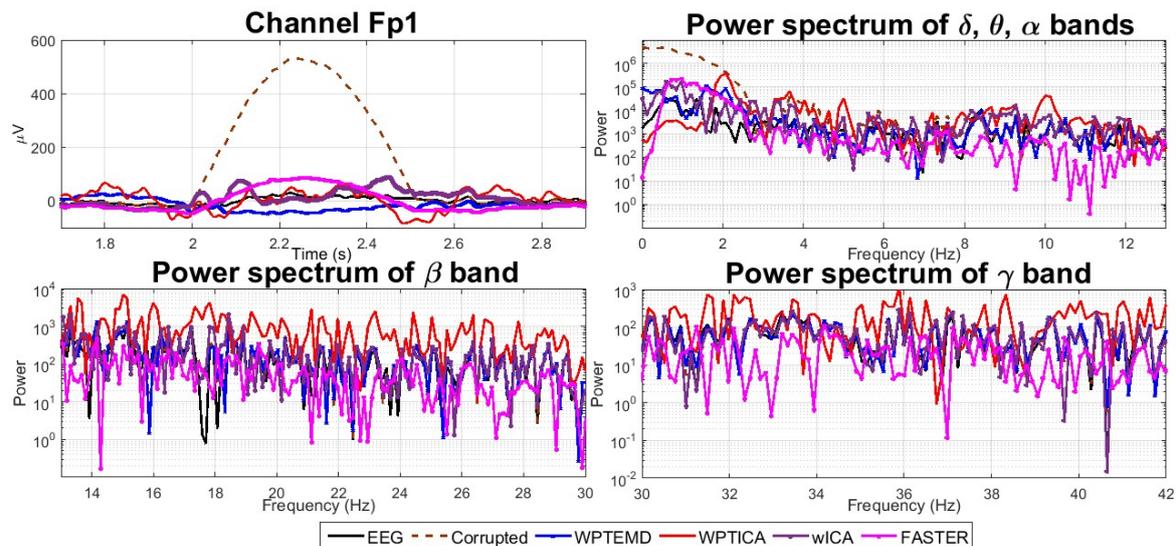

Figure 7: Semi-simulated eye-blink artifact in the channel Fp1 in time domain, and in frequency domain in each EEG frequency band ($\delta$, $\theta$, $\alpha$, $\beta$ and $\gamma$).

Overall, the exploration with the semi-simulated data shows that WPTEMD outperforms the rest of the methods (both WPTICA and the two benchmark algorithms) in case of heavily contaminated data (frontal channels and high $\lambda$ in the lower frequency bands where eye-blink is contaminating the EEG).

### 3.2. Performance Evaluation with the Real Pervasive EEG Artifacts

The previous subsection shows the analysis of semi-simulated EEG, here we explore eight types of real artifact examples commonly encountered in pervasive EEG system during natural body movement. Some of the artifacts, like eye-blink or talking, affect only a few of the channels but the stronger artifacts like the yaw, pitch and roll head movement corrupt most of the electrodes at the same time. In Table 1, we have shown a summary of the most corrupted channel for each type of artifact in each band using all trials and subjects (3×10).

Due to the unavailability of the ground truth signal (unlike the semi-simulated case reported in earlier subsection), here we report only the *ASR* based and scalp topography based assessment of the cleaning performance of the two hybrid algorithms.

Table 1
Channel with the highest power for each artifact and for each band across all trials and all subjects

| Artifact type | $\delta$ | $\theta$ | $\alpha$ | $\beta$ | $\gamma$ |
|---|---|---|---|---|---|
| Eye-blink | $F_{p2}$ | $F_{p1}$ | $F_{p1}$ | $F_{p1}$ | $T_8$ |
| Hello | $F_{p2}$ | $F_{p2}$ | $F_z$ | $F_4$ | $T_8$ |
| Chewing | $C_4$ | $F_7$ | $F_7$ | $F_4$ | $F_4$ |
| Left hand | $F_{p1}$ | $F_{p1}$ | $F_{p1}$ | $F_3$ | $T_8$ |
| Right hand | $F_{p1}$ | $F_{p1}$ | $T_7$ | $F_3$ | $F_{p1}$ |
| Left right head | $T_8$ | $F_{p2}$ | $F_{p2}$ | $F_4$ | $F_{p1}$ |
| Down up head | $P_4$ | $P_z$ | $P_z$ | $O_2$ | $O_2$ |
| Shaking head | $C_4$ | $C_4$ | $C_4$ | $C_4$ | $F_4$ |

### 3.2.1. Comparison of Algorithms Using ASR Criterion

From the exploration with the semi-simulated data it is clear that the *ASR* criterion could be considered as an effective quantitative metric for evaluating the performance of the algorithms for cleaning artifacts. The *ASR* parameter has been calculated for real data across multiple subjects and trials and for each type of artifact in each band to investigate: (a) which channels and (b) bands are mostly modified by each of the algorithms and (c) also to quantify the variability of the artifact suppression performance across multiple trials and subjects. The electrodes with high *ASR* are the ones which have been modified by the algorithm to a higher extent.

Figure 8 shows the performance of WPTEMD and WPTICA in terms of *ASR* and the corresponding power for each electrode. As evident from **Error! Reference source not found.**, the *ASR* values in $\alpha$, $\beta$ and $\gamma$ bands have a lower range compared to the $\delta$ and $\theta$ bands, suggesting that the high frequency bands have not been severely modified by the algorithms. The power in $\delta$, $\theta$ and $\alpha$ bands is higher in the channels Fp1 and Fp2 and the corresponding *ASR* values for both algorithms are higher than the rest of the channels. This indicates that the algorithms have modified mainly these two electrodes: in $\delta$ and $\theta$ bands the performances are similar, whereas in $\alpha$ band the negative *ASR* indicates that the WPTICA increased the power of the reconstructed signal. In $\beta$ and $\gamma$ bands, even though the power is similar in all electrodes of the corrupted EEG, WPTICA increases the power spectrum of the frontal channels ($ASR < 0$).

In the higher frequency bands ($\beta$ and $\gamma$), similar results are found for the other types of artifact (please refer to the supplementary material), where the median of the *ASR* is very low, indicating that the power of the 'clean' signal is nearly equal to the corrupted EEG. Hence, the algorithms have not modified the higher frequency bands.

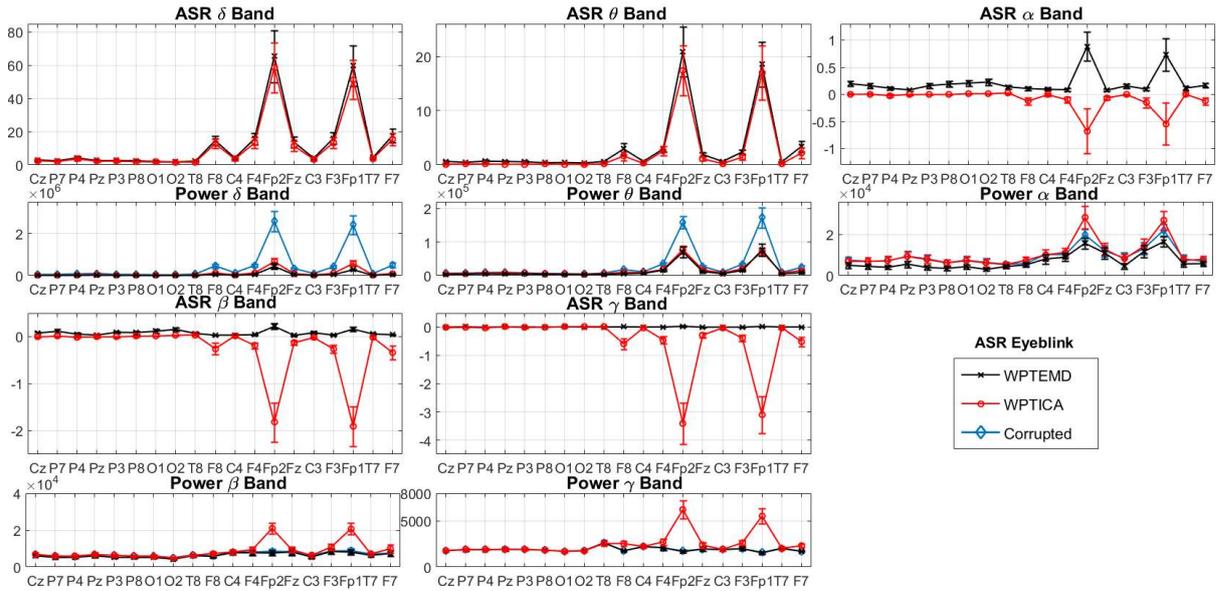

Figure 8: *ASR* and power of real eye-blink EEG data across all the subjects and trials for each EEG band. WPTEMD consistently decreases (in $\theta$, $\delta$ and $\alpha$ bands) or preserves ($\beta$ and $\gamma$ bands) the power of the corrupted EEG recordings. WPTICA shows negative *ASR*: it increases the power of the corrupted EEG in the higher frequency bands ($\alpha$, $\beta$ and $\gamma$).

### 3.2.2.  *Real Pervasive EEG Artifact Data: Scalp Topography Analysis*

An alternative way to investigate the algorithms' performance is to show how the scalp topography of the EEG varies when the signal is affected by a specific type of artifact and when each of the proposed algorithms is applied. The scalp topography shows the mean power across all subjects for each EEG signal (i.e. Resting, Corrupted, WPT, WPTEMD and WPTICA), for all the frequency ranges (*All bands*, i.e. from $\delta$ to $\gamma$) and for each band separately (i.e. $\delta$, $\theta$, $\alpha$, $\beta$ and $\gamma$). The power shown at each scalp is normalized within each band and across the different signals (i.e. Resting, Corrupted, WPT, WPTEMD and WPTICA). The distribution of the power over the scalp was shown for each band to highlight how the algorithms affect each frequency band. The power spectrum of each electrode has been estimated through the Fast Fourier Transform magnitude squared. Such power is also normalized between zero and one for five different cases: Resting, Corrupted, WPT, WPTEMD and WPTICA. The resting signal is shown here because its parameters (i.e. entropy and standard deviation) are used to calculate the *J* parameter in the WPTEMD algorithm, but none of the algorithms modifies it. However, the resting EEG appears different in each artifact case because of the normalization across the different signals (i.e. Resting, Corrupted, WPT, WPTEMD and WPTICA) within each frequency band (i.e. *All bands*, $\delta$, $\theta$, $\alpha$, $\beta$ and $\gamma$).

Figure 9 (a) shows the scalp topographical maps during the eye-blink artifact. From the scalp topography across all bands (i.e. *All bands*) of the corrupted signal, it is evident that only the frontal region shows high power. The same pattern is present in all the frequency bands except $\beta$ and $\gamma$, suggesting that the artifact affects the EEG power in the frequency range up to 13 Hz. The WPT is capable of reducing most of the power in the frontal electrodes only in the $\delta$ band, while the WPTEMD technique reduces the power in all the affected frequency bands (i.e. $\delta$, $\theta$ and $\alpha$). WPTICA, instead, reduces the power of the frontal electrodes in both $\delta$ and $\theta$ bands, but it increases the power in the rest of the bands.

Figure 9 (b) shows the topographical maps related to the yaw movement of the head. The corrupted signal is characterized by high power in the left temporal and occipital electrodes, as visible in the scalp topography-*All bands* and the same pattern is present in the $\delta$ band. This suggests that $\delta$ is the most corrupted band and this contamination is fully removed by all the algorithms. The rest of the bands are characterized by a common pattern of high power across the head. Both WPTEMD and WPTICA decrease this power across all the bands (from $\theta$ to $\gamma$).

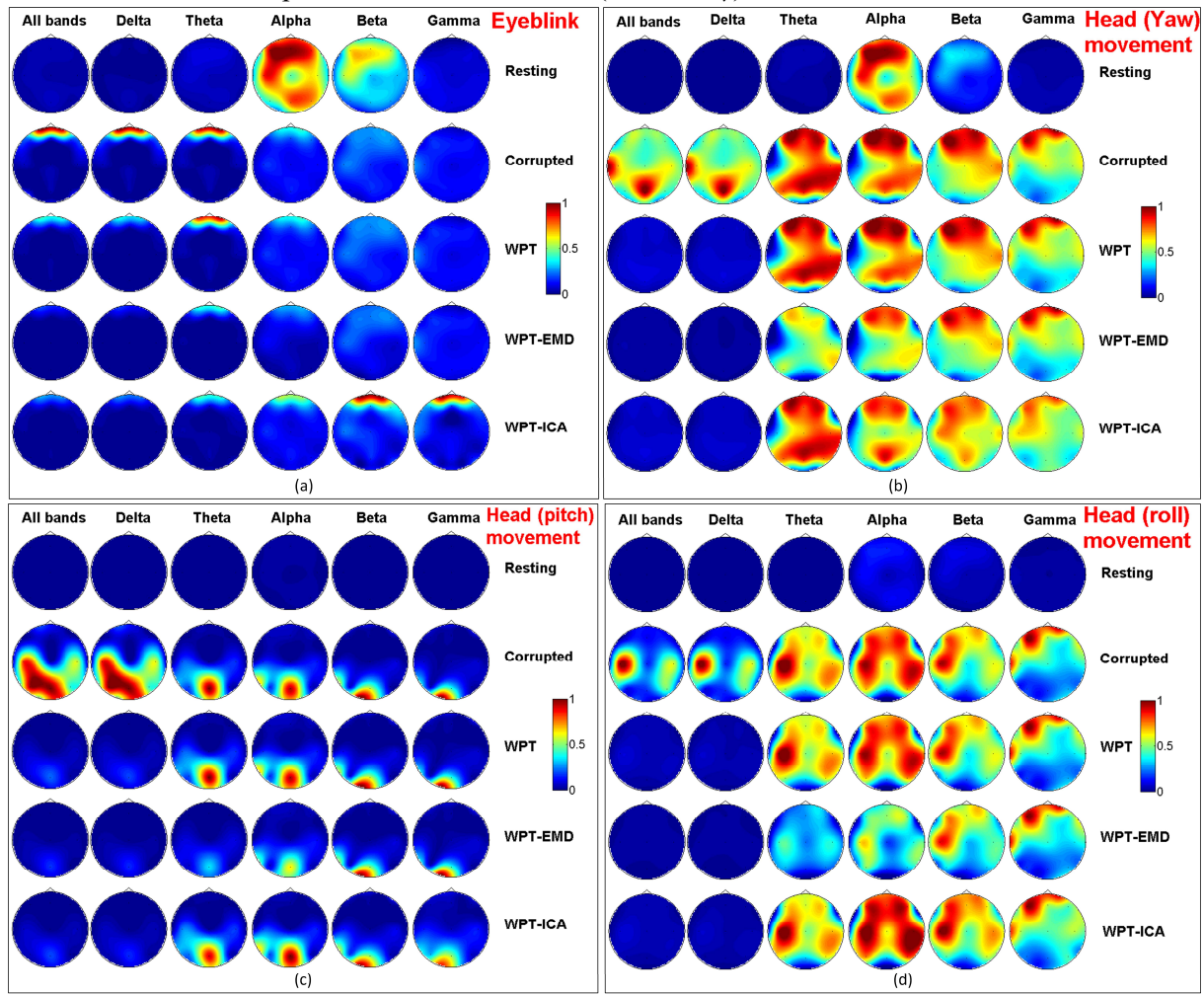

Figure 9: Scalp topography of EEG recordings contaminated by different artifacts (i.e. (a) eye-blink, (b) head movement in yaw, (c) pitch and (d) roll), across different subjects and trials. The power shown at each scalp has been normalized within each band and across the different signals (i.e. Resting, Corrupted, WPT, WPTEMD and WPTICA). The high power in the scalp in *All bands* is present in $\delta$ band, suggesting that the artifact contamination takes place mainly in $\delta$ band. Among all the techniques, WPTEMD reduces the localized high power in a higher extent.

The head (pitch) movement causes high power in the occipital electrodes, as visible in the scalp topography-*All bands* in Figure 9 (c). The occipital region shows high power in all the frequency bands (from $\delta$ to $\gamma$), suggesting that this artifact affects all of them. The WPTEMD algorithm is the most capable one for reducing the high power in all the bands, except the $\beta$ and $\gamma$ bands which show the same pattern in the signal even after the application of all algorithms. It is worth noting that in case of WPT the overall (i.e. *All bands*) power in the occipital channels is reduced following the reduction of the power in $\delta$ band only.

Head (roll) movement affects the temporal lobes, as shown in the scalp topography-*All bands* in Figure 9 (d). A similar topography pattern is visible in the $\delta$ band, while the other bands show high power also in the frontal region. All the algorithms are capable of removing the artifact in the $\delta$ band, due to the action of the WPT technique. The WPTEMD is also capable of reducing the power in the $\theta$ and $\alpha$ bands, whereas WPTICA increases the power in $\alpha$ and $\beta$ bands.

Figure 10 (a) shows that the highest power is located in the frontal, left temporal and occipital regions for the left hand movement artifact. The high power in the scalp topography-*All bands* is also visible in the $\delta$ and $\theta$ bands, suggesting that the artifact might affect the power in the frequency range up to 8 Hz, which is mostly suppressed by the WPTEMD algorithm. Unlike the WPTEMD, the scalp topography in WPT and WPTICA still contains high power in the frontal region in the $\theta$ band.

The scalp topography- *All bands* from data recorded while the subject was performing right hand movement shows high power in the frontal region, as can be seen in Figure 10 (b). A similar pattern is also visible in the $\delta$, $\theta$ and $\beta$ bands. WPT decreases the power in *All bands* and in $\delta$ band. Additionally, WPTEMD decreases the power of the frontal electrodes also in the $\theta$ and $\beta$ bands. WPTICA decreases the power of $\theta$ band but does not modify $\beta$ and $\gamma$ bands.

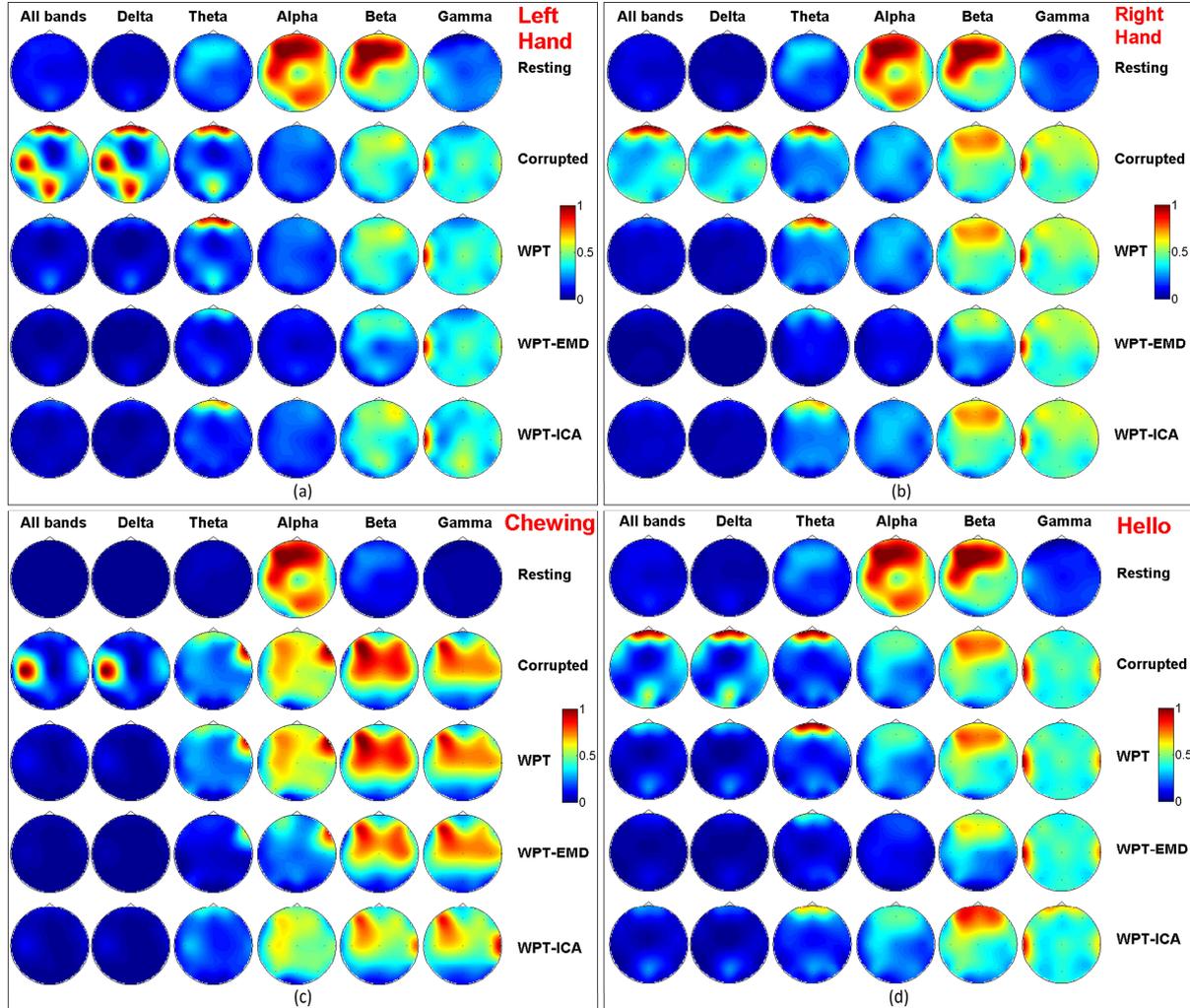

Figure 10: Scalp topography of EEG recordings contaminated by different artifacts (i.e. hand movements (a, b), chewing (c) and talking (d)), across different subjects and trials. The power shown at each scalp has been normalized within each band and across the different signals (i.e. Resting, Corrupted, WPT, WPTEMD and WPTICA). The high power in the scalp in *All bands* is present in $\delta$ band, suggesting that the artifact contamination takes place mainly in $\delta$ band. Among all the techniques, WPTEMD reduces the localized high power in a higher extent in most of the artifacts. WPTICA in some cases increases the power of the reconstructed signal.

Figure 10 (c) shows that the left temporal lobe has the highest power in case of chewing, as visible in the scalp topography-*All bands*. The same pattern is detectable in the $\delta$ band and it is reduced by all the techniques. The high power characterizing the rest of the bands is reduced in a higher extent by the WPTEMD algorithm compared to the other variants. WPTICA decreases the power in all the bands but it also increases the power of the right temporal lobe in $\gamma$ band.

Talking affects the frontal and occipital regions, as shown in the scalp topography-*All bands* in Figure 10 (d). These high power patterns are also present in the $\delta$ band, in a greater extent, and in all the other bands except the $\gamma$ band. The highest power suppression is achieved by the WPTEMD technique in all the bands except the $\gamma$, while WPTICA increases the power in the frontal electrodes in $\beta$ and $\gamma$ bands.

### 3.2.3. Real Pervasive EEG Artifact data: Time and Frequency Domain Performance Analysis

Here we explore the time and frequency domain characteristics of the single subject-single trial EEG, before and after the application of the artifact suppression algorithms. In order to highlight the specific modifications of the frequency spectrum in different EEG bands of interest, we have shown the analysis on lower ($\delta$, $\theta$ and $\alpha$) and higher ($\beta$ and $\gamma$) bands separately. The power spectrum has been estimated here through Fast Fourier transform magnitude squared and the average power in each band has been calculated integrating the Power Spectral Density (PSD) curve over the frequency band of interest. The suppression of artifacts in time domain is also evident by the reduction of its amplitude in one of the most affected channels.

Figure 11 shows an example of the eye-blink artifact in time and frequency domain. The spikes related to the artifacts are reduced in the case of WPTEMD, but not for the WPT and the WPTICA. In the low frequency bands (i.e. $\delta$, $\theta$ and $\alpha$), the power spectrum is modified mostly below 6 Hz, since the WPT acts as a band stop filter between 3 and 6 Hz. The WPTEMD preserves the power spectrum in the rest of the frequency bands, while the WPTICA distorts the power spectrum almost in all the bands and increases the power in $\beta$ and $\gamma$ bands.

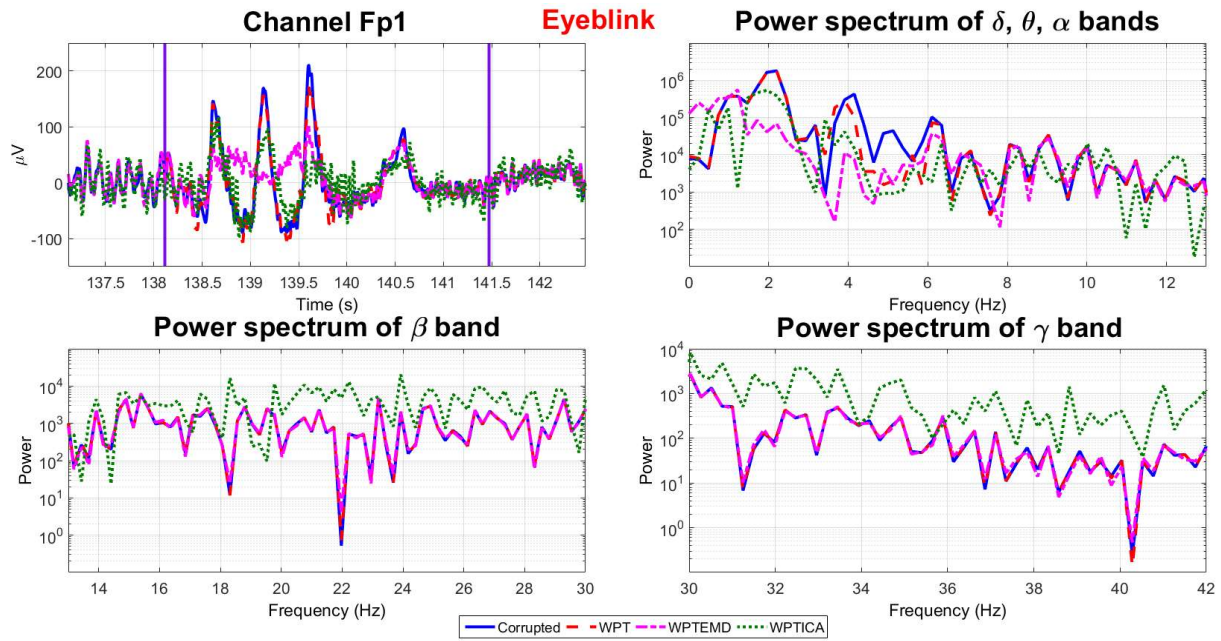

Figure 11: EEG data in time and frequency domain contaminated by the eye-blink artifact in channel Fp1. WPTEMD is the most capable of reducing the high amplitude spikes and preserves the power spectrum of the EEG signal in the higher frequency bands.

Unlike the eye-blink, the head (yaw) movement generates artifacts which affect almost all the channels. Figure 12 shows an example of head movement artifact in the frontal electrode Fp2. In time domain, the WPT and the WPTICA signals overlap, while the WPTEMD reduces the spikes in the time intervals 131.5−132 s and 134.5−135.5 s. Below 2 Hz the power spectrum of the corrupted signal is modified by the WPT, but it is preserved in the rest of the bands. The WPTEMD modified the power spectrum up to 10 Hz. The rest of the bands have not been affected by any of the techniques.

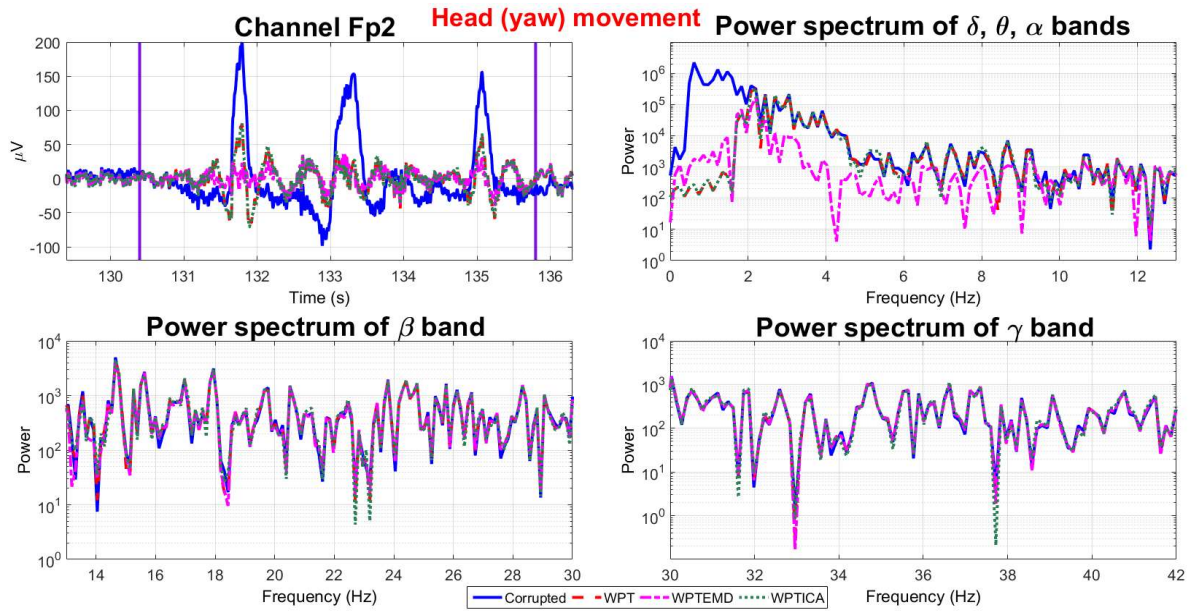

Figure 12: EEG data in time and frequency domain contaminated by the left right (yaw) head movement in channel Fp2. WPTEMD is the most capable of reducing the high amplitude spikes and reduces the power spectrum of the EEG signal in the lower frequency bands.

Similar results both in time and frequency domains have been found in the EEG acquired during the head (pitch) movement, as shown in Figure 13. In time domain, the spikes are suppressed by all the applied techniques. The power spectrum only in the low frequency bands is affected by all the algorithms. The WPT removes the low frequency components (below 2 Hz) and it is overlapped with the spectrum related to WPTICA, while the WPTEMD algorithm modified the spectrum up to 6 Hz.

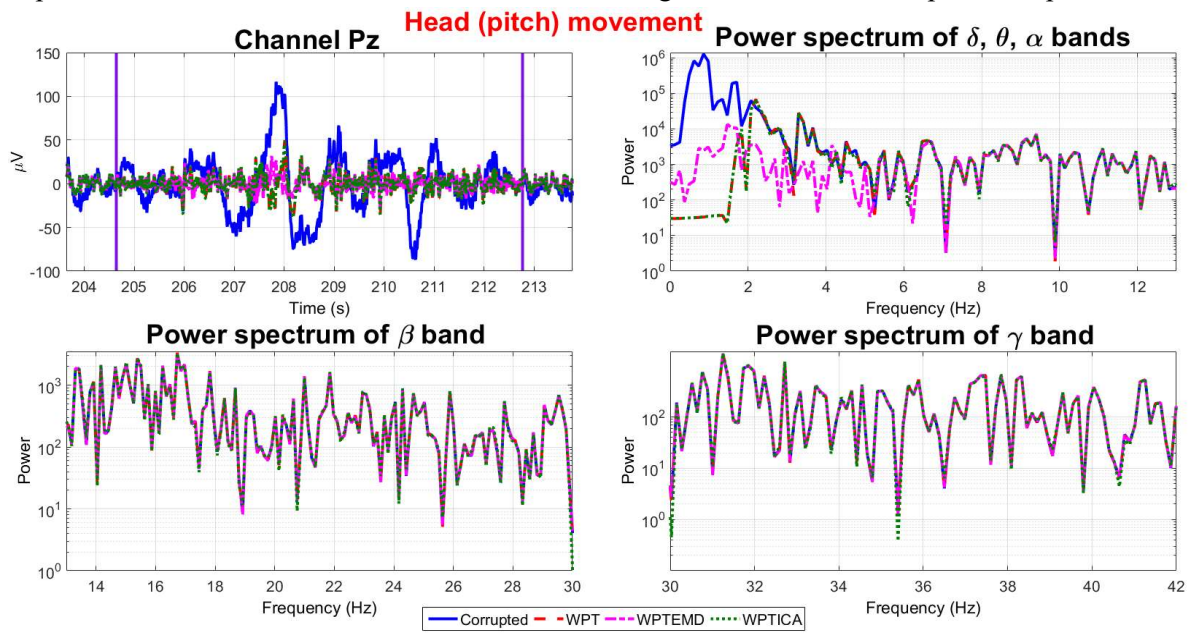

Figure 13: EEG data in time and frequency domain contaminated by the down up (pitch) head movement in channel Pz. The high amplitude spikes are reduced in a higher extent by the WPTEMD which modifies the power spectrum of the EEG signal up to 7 Hz.

When considering shaking the head sidewise (roll), the amplitude of EEG is found to be much higher compared to the artifacts previously described (10 times bigger than the clean EEG) as shown in Figure 14. The spikes are reduced by all the techniques, but the WPTEMD showed the best results, even though some slow and big oscillations are still visible. The $\delta$, $\theta$ and $\alpha$ bands contain the highest power spectrum, which is modified only below 2 Hz by the WPT and WPTICA and up to 10 Hz by the WPTEMD.

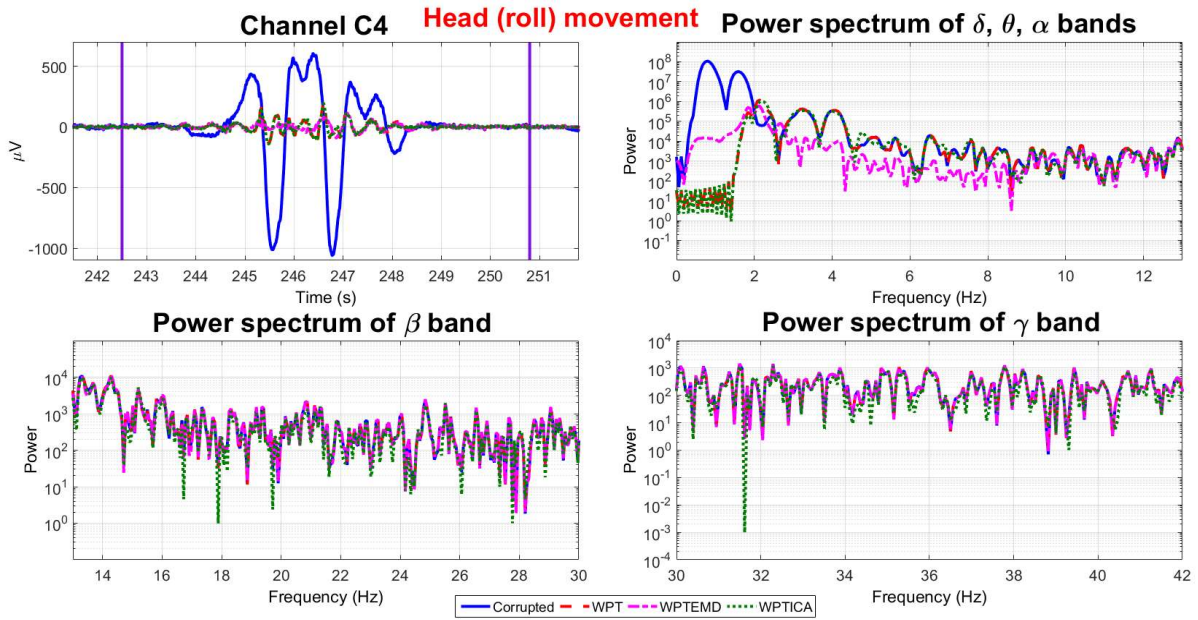

Figure 14: EEG data in time and frequency domain contaminated by the shaking head (roll) movement in channel C4. The high amplitude spikes are reduced in a higher extent by the WPTEMD which modifies the power spectrum of the EEG signal up to 9 Hz.

An example of the left hand-movement is shown in Figure 15: the high amplitude oscillations in the time interval 322-324 s are significantly reduced and the best performance is given by the WPTEMD. Only the low frequency spectrum is modified: the WPT and WPTICA altered the spectrum below 2 Hz, while the WPTEMD modified the frequency components of the signal up to 16 Hz.

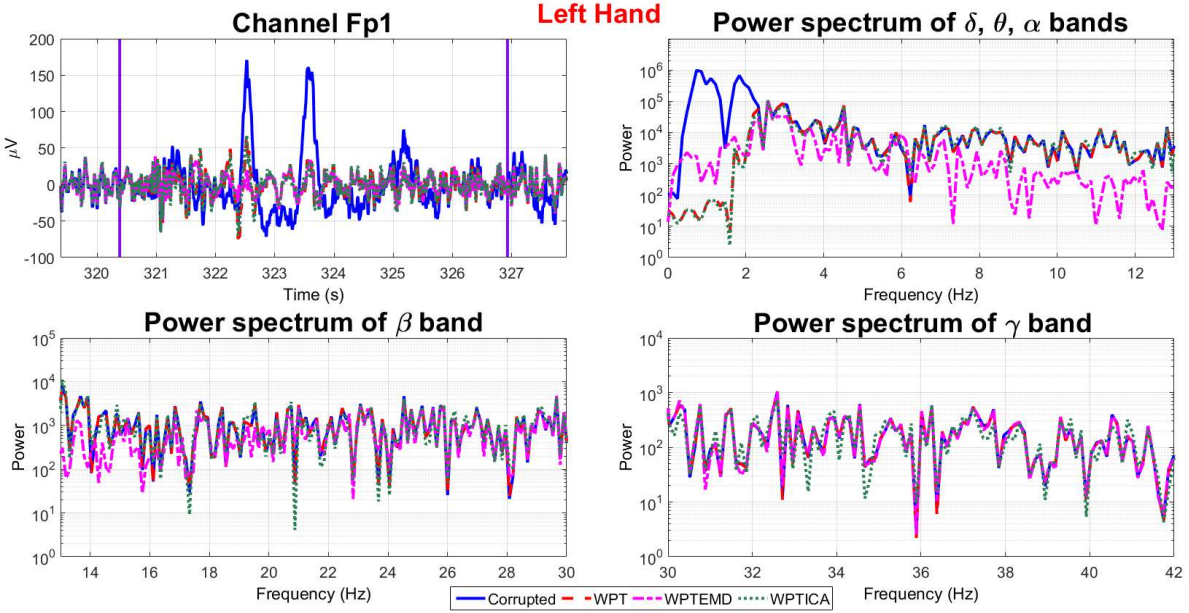

Figure 15: EEG data in time and frequency domain contaminated by the left hand movement in channel Fp1. The high amplitude spikes are reduced in a higher extent by the WPTEMD which modifies the power spectrum of the EEG signal up to 17 Hz.

Figure 16 shows that the WPTEMD is better capable of reducing the large oscillations caused by the right hand movement. In frequency domain, the spectrum of the corrupted signal is modified only up to 10 Hz in case of WPTEMD and below 3 Hz for the other algorithms.

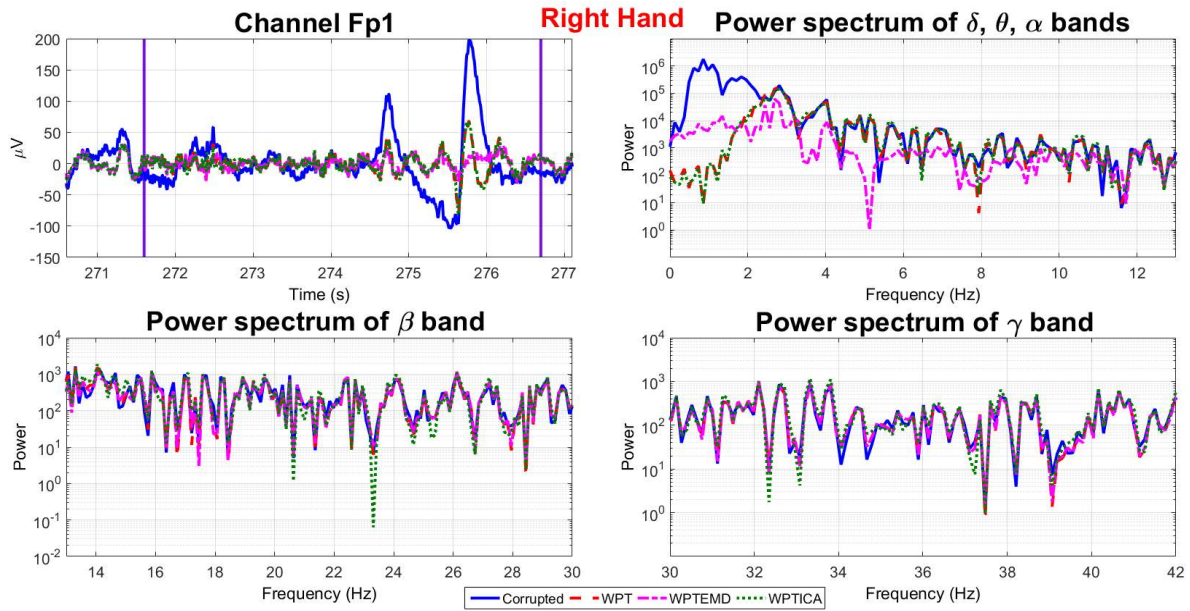

Figure 16: EEG data in time and frequency domain contaminated by the right hand movement in channel Fp1. The high amplitude spikes are reduced in a higher extent by the WPTEMD, as visible around second 276.

The frontal and temporal electrodes are affected by the artifacts caused by the muscles activity during chewing. The high amplitude oscillations are better reduced by the WPTEMD as shown in Figure 17. The power spectrum is slightly modified by the WPT and WPTICA below 2 Hz, while is reduced up to 6 Hz in the case of WPTEMD.

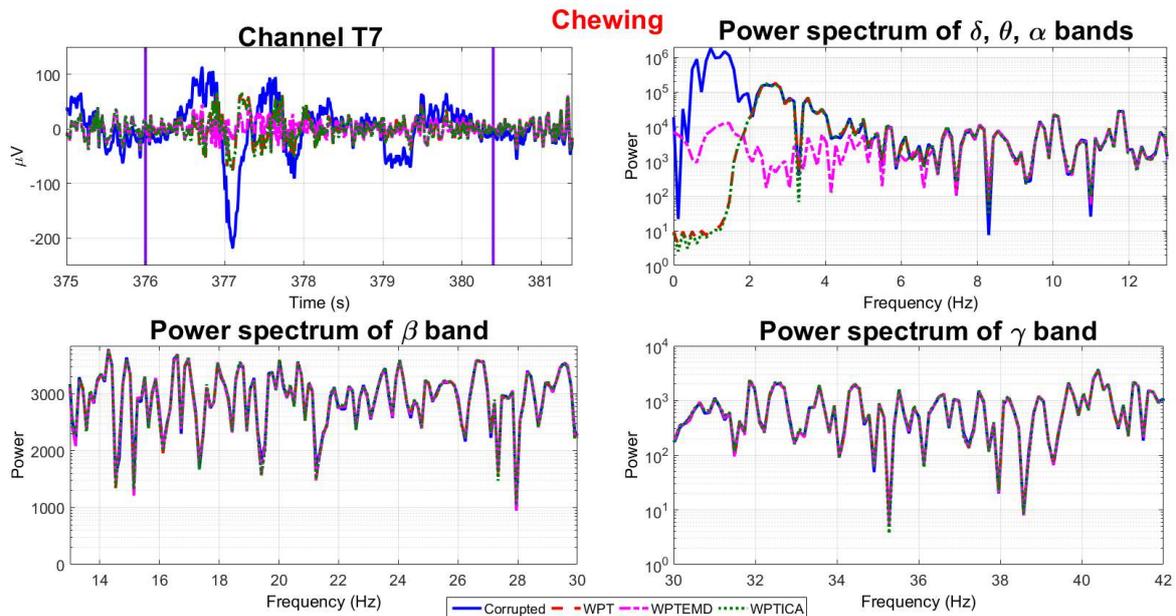

Figure 17: EEG data in time and frequency domain contaminated by the chewing in channel T7. The high amplitude spikes are reduced in a higher extent by the WPTEMD, as visible around second 377.

The last type of artifact investigated here was caused while the subject was asked to pronounce the word 'hello' to evaluate the effect of speaking on the acquired EEG. The frontal electrodes are likely to capture the effect of facial muscle activity. Figure 18 shows that the high amplitude oscillations in the signal are better reduced by the WPTEMD technique. The WPT acts as a high-pass filter with a cutoff frequency around 2 Hz. The spectrum is modified in all the frequency bands by the WPTICA. The WPT and WPTEMD techniques alter only the spectrum in the $\delta$, $\theta$ and $\alpha$ bands.

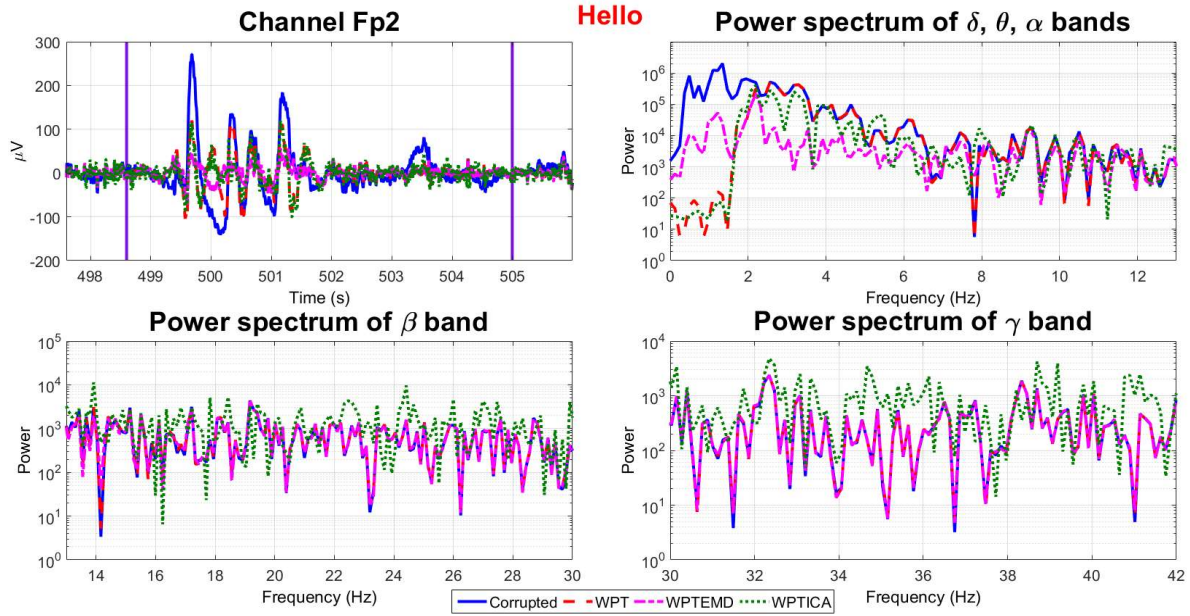

Figure 18: EEG data in time and frequency domain contaminated by talking in channel T7. The high amplitude spikes are reduced in a higher extent by the WPTEMD, which preserves the power spectrum in the higher frequency bands.

From the above exploration it is evident that the two proposed hybrid algorithms outperforms the benchmark algorithms and the basic constituents – WPT, ICA and EMD for both the semi-simulated and real pervasive EEG. We have validated our results on multiple subject and multiple trials of each of the eight classes of artifacts. The results confirm that WPTEMD performs better than WPTICA in case of heavily corrupted data [29] when dealing with low number of electrodes (i.e. 19 channels) and motion artifacts. Unlike WPTICA, WPTEMD requires the availability of the EEG recording with no artifact to calculate the $J$ parameter. However, we do not consider this requirement as a limitation of the method since it is a common practice to include resting EEG with closed and open eyes during a recording session [51]. We have run the algorithms with Matlab (release R2013b) on a Windows 7 PC with Intel(R) Core i7-2600 CPU @ 3.4 GHz processor with 16 GB RAM. The run time for the semi-simulated data is in the order of 9 seconds and is of similar order in all the proposed methods and benchmarks. The detailed statistical analysis reported in the previous sections are provided in the supplementary material for brevity, in particular – 1) surface plots of $J$ parameter with channel and weight, 2) $J$ parameters of different IMFs across channels, 3) variation of *ASR* and power in different bands and channels and 4) single subject scalp topography across all bands, all repeated for other artifact types apart from eye-blink.

## 4. Conclusions

In this paper we explore the performance of two hybrid algorithms – WPTEMD and WPTICA – for suppressing motion related artifacts when *a priori* knowledge about the characteristics of the artifacts is not available; more specifically in the case of pervasive EEG recording. The exploration was first carried out with semi-simulated data where the above mentioned algorithms were compared with state-of-the-art artifact separation algorithms like wICA and FASTER. The performances have been analyzed by comparing the reconstructed signals after processing the artifacts with the available artifact-free ground truth signal. Later we explored the performance with experimentally acquired data corrupted with eight types of motion artifacts across multiple subjects and trials using 19-channel commercially available pervasive EEG system called Enobio.

The exploration with semi-simulated data shows that WPTEMD algorithm outperforms all the other variants under investigation, in terms of *RMSE* for highly corrupted channels (49.53% and 70.4% performance improvement compared to wICA and FASTER respectively) and maintains the power spectrum of different frequency bands close to the original semi-simulated signals. When analyzed with the real data, the performances of the algorithms showed similar results: for highly corrupted channels WPTEMD consistently shows better performance than WPTICA in all the frequency bands and reduces the high-amplitude oscillations in the time-domain signal. This result holds true for all the eight different types of artifacts we have explored. Therefore our exploration indicates that for a pervasive EEG recording scenario WPTEMD could be the best choice for suppressing unwanted motion artifacts during natural movement of the subject.

## Acknowledgement:

This work was supported by FP7 EU funded MICHELANGELO project, Grant Agreement #288241. URL: www.michelangelo-project.eu/.

## Reference:

[1]   S. Sanei and J. A. Chambers, *EEG signal processing*. John Wiley & Sons, 2008.
[2]   B. W. McMenamin, A. J. Shackman, L. L. Greischar, and R. J. Davidson, "Electromyogenic artifacts and electroencephalographic inferences revisited," *NeuroImage*, vol. 54, no. 1, pp. 4–9, 2011.
[3]   M. Crespo-Garcia, M. Atienza, and J. L. Cantero, "Muscle artifact removal from human sleep EEG by using independent component analysis," *Annals of Biomedical Engineering*, vol. 36, no. 3, pp. 467–475, 2008.
[4]   C. A. Joyce, I. F. Gorodnitsky, and M. Kutas, "Automatic removal of eye movement and blink artifacts from EEG data using blind component separation," *Psychophysiology*, vol. 41, no. 2, pp. 313–325, 2004.
[5]   K. Ting, P. Fung, C. Chang, and F. Chan, "Automatic correction of artifact from single-trial event-related potentials by blind source separation using second order statistics only," *Medical Engineering & Physics*, vol. 28, no. 8, pp. 780–794, 2006.
[6]   "Enobio." [Online]. Available: http://www.neuroelectrics.com/products/enobio/enobio-20/
[7]   Y. M. Chi, T.-P. Jung, and G. Cauwenberghs, "Dry-contact and noncontact biopotential electrodes: methodological review," *Biomedical Engineering, IEEE Reviews in*, vol. 3, pp. 106–119, 2010.
[8]   B. Hu, C. Mao, W. Campbell, P. Moore, L. Liu, and G. Zhao, "A pervasive EEG-based biometric system," in *Proceedings of 2011 International Workshop on Ubiquitous Affective Awareness and Intelligent Interaction*, 2011, pp. 17–24.
[9]   L. Sörnmo and P. Laguna, Bioelectrical signal processing in cardiac and neurological applications. Academic Press, 2005.
[10]  H. Nolan, R. Whelan, and R. Reilly, "FASTER: fully automated statistical thresholding for EEG artifact rejection," *Journal of neuroscience methods*, vol. 192, no. 1, pp. 152–162, 2010.
[11]  J. A. Urigüen and B. Garcia-Zapirain, "EEG artifact removal—state-of-the-art and guidelines," *Journal of neural engineering*, vol. 12, no. 3, p. 031001, 2015.
[12]  N. P. Castellanos and V. A. Makarov, "Recovering EEG brain signals: artifact suppression with wavelet enhanced independent component analysis," *Journal of Neuroscience Methods*, vol. 158, no. 2, pp. 300–312, 2006.
[13]  V. Bono, W. Jamal, S. Das, and K. Maharatna, "Artifact reduction in multichannel pervasive EEG using hybrid WPT-ICA and WPT-EMD signal decomposition techniques," in *Acoustics, Speech and Signal Processing (ICASSP), 2014 IEEE International Conference on*, 2014, pp. 5864 – 5868.
[14]  S. Narasimhan and D. N. Dutt, "Application of LMS adaptive predictive filtering for muscle artifact (noise) cancellation from EEG signals," *Computers & Electrical Engineering*, vol. 22, no. 1, pp. 13–30, 1996.
[15]  M. Fatourechi, A. Bashashati, R. K. Ward, and G. E. Birch, "EMG and EOG artifacts in brain computer interface systems: A survey," *Clinical Neurophysiology*, vol. 118, no. 3, pp. 480–494, 2007.
[16]  D. Moretti, F. Babiloni, F. Carducci, F. Cincotti, E. Remondini, P. Rossini, S. Salinari, and C. Babiloni, "Computerized processing of EEG-EOG-EMG artifacts for multi-centric studies in EEG oscillations and event-related potentials," *International Journal of Psychophysiology*, vol. 47, no. 3, pp. 199–216, 2003.
[17]  K. T. Sweeney, T. E. Ward, and S. F. McLoone, "Artifact removal in physiological signals—Practices and possibilities," *Information Technology in Biomedicine, IEEE Transactions on*, vol. 16, no. 3, pp. 488–500, 2012.
[18]  T.-P. Jung, S. Makeig, C. Humphries, T.-W. Lee, M. J. Mckeown, V. Iragui, and T. J. Sejnowski, "Removing electroencephalographic artifacts by blind source separation," *Psychophysiology*, vol. 37, no. 2, pp. 163–178, 2000.

[19] W. De Clercq, A. Vergult, B. Vanrumste, W. Van Paesschen, and S. Van Huffel, "Canonical correlation analysis applied to remove muscle artifacts from the electroencephalogram," *Biomedical Engineering, IEEE Transactions on*, vol. 53, no. 12, pp. 2583–2587, 2006.

[20] H. Hallez, M. De Vos, B. Vanrumste, P. Van Hese, S. Assecondi, K. Van Laere, P. Dupont, W. Van Paesschen, S. Van Huffel, and I. Lemahieu, "Removing muscle and eye artifacts using blind source separation techniques in ictal EEG source imaging," *Clinical Neurophysiology*, vol. 120, no. 7, pp. 1262–1272, 2009.

[21] A. Vergult, W. De Clercq, A. Palmini, B. Vanrumste, P. Dupont, S. Van Huffel, and W. Van Paesschen, "Improving the interpretation of ictal scalp EEG: BSS-CCA algorithm for muscle artifact removal," *Epilepsia*, vol. 48, no. 5, pp. 950–958, 2007.

[22] F.-B. Vialatte, J. Solé-Casals, and A. Cichocki, "EEG windowed statistical wavelet scoring for evaluation and discrimination of muscular artifacts," *Physiological Measurement*, vol. 29, no. 12, p. 1435, 2008.

[23] M. T. Akhtar, W. Mitsuhashi, and C. J. James, "Employing spatially constrained ICA and wavelet denoising, for automatic removal of artifacts from multichannel EEG data," *Signal Processing*, vol. 92, no. 2, pp. 401–416, 2012.

[24] M. Zima, P. Tichavský, K. Paul, and V. Krajča, "Robust removal of short-duration artifacts in long neonatal EEG recordings using wavelet-enhanced ICA and adaptive combining of tentative reconstructions," *Physiological Measurement*, vol. 33, no. 8, p. N39, 2012.

[25] L. Zou, S. Xu, Z. Ma, J. Lu, and W. Su, "Automatic Removal of Artifacts from Attention Deficit Hyperactivity Disorder Electroencephalograms Based on Independent Component Analysis," *Cognitive Computation*, vol. 5, no. 2, pp. 225–233, 2013.

[26] B. Mijović, M. De Vos, I. Gligorijević, J. Taelman, and S. Van Huffel, "Source separation from single-channel recordings by combining empirical-mode decomposition and independent component analysis," *Biomedical Engineering, IEEE Transactions on*, vol. 57, no. 9, pp. 2188–2196, 2010.

[27] K. T. Sweeney, H. Ayaz, T. E. Ward, M. Izzetoglu, S. F. McLoone, and B. Onaral, "A methodology for validating artifact removal techniques for physiological signals," *Information Technology in Biomedicine, IEEE Transactions on*, vol. 16, no. 5, pp. 918–926, 2012.

[28] K. T. Sweeney, S. F. McLoone, and T. E. Ward, "The use of ensemble empirical mode decomposition with canonical correlation analysis as a novel artifact removal technique," *Biomedical Engineering, IEEE Transactions on*, vol. 60, no. 1, pp. 97–105, 2013.

[29] D. Safieddine, A. Kachenoura, L. Albera, G. Birot, A. Karfoul, A. Pasnicu, A. Biraben, F. Wendling, L. Senhadji, and I. Merlet, "Removal of muscle artifact from EEG data: comparison between stochastic (ICA and CCA) and deterministic (EMD and wavelet-based) approaches," *EURASIP Journal on Advances in Signal Processing*, vol. 2012, no. 1, pp. 1–15, 2012.

[30] C. W. Hesse and C. J. James, "On semi-blind source separation using spatial constraints with applications in EEG analysis," *Biomedical Engineering, IEEE Transactions on*, vol. 53, no. 12, pp. 2525–2534, 2006.

[31] C. W. Hesse and C. J. James, "The FastICA algorithm with spatial constraints," *Signal Processing Letters, IEEE*, vol. 12, no. 11, pp. 792–795, 2005.

[32] F. C. Robertson, T. S. Douglas, and E. M. Meintjes, "Motion artifact removal for functional near infrared spectroscopy: a comparison of methods," *Biomedical Engineering, IEEE Transactions on*, vol. 57, no. 6, pp. 1377–1387, 2010.

[33] B. Walczak and D. Massart, "Noise suppression and signal compression using the wavelet packet transform," *Chemometrics and Intelligent Laboratory Systems*, vol. 36, no. 2, pp. 81–94, 1997.

[34] L. Fraiwan, N. Khaswaneh, and K. Y. Lweesy, "Automatic sleep stage scoring with wavelet packets based on single EEG recording," *World Academy of Science, Engineering and Technology*, vol. 54, pp. 485–488, 2009.

[35] L. Yuan, B. Yang, S. Ma, and B. Cen, "Combination of wavelet packet transform and Hilbert-Huang transform for recognition of continuous EEG in BCIs," in *Computer Science and Information Technology, 2009. ICCSIT 2009. 2nd IEEE International Conference on*, 2009, pp. 594–599.

[36] D. Wang, D. Miao, and C. Xie, "Best basis-based wavelet packet entropy feature extraction and hierarchical EEG classification for epileptic detection," *Expert Systems with Applications*, vol. 38, no. 11, pp. 14314–14320, 2011.

[37] A. Hyvärinen and E. Oja, "Independent component analysis: algorithms and applications," *Neural Networks*, vol. 13, no. 4, pp. 411–430, 2000.

[38] K. A. Glass, G. A. Frishkoff, R. M. Frank, C. Davey, J. Dien, A. D. Malony, and D. M. Tucker, "A framework for evaluating ICA methods of artifact removal from multichannel EEG," in *Independent Component Analysis and Blind Signal Separation*, Springer, 2004, pp. 1033–1040.

[39] S. Gordon, V. Lawhern, A. Passaro, and K. McDowell, "Informed decomposition of electroencephalographic data," *Journal of neuroscience methods*, vol. 256, pp. 41–55, 2015.

[40] A. Hyvarinen, "Fast and robust fixed-point algorithms for independent component analysis," *Neural Networks, IEEE Transactions on*, vol. 10, no. 3, pp. 626–634, 1999.

[41] Y. Li, Z. Ma, W. Lu, and Y. Li, "Automatic removal of the eye blink artifact from EEG using an ICA-based template matching approach," *Physiological Measurement*, vol. 27, no. 4, p. 425, 2006.

[42] N. E. Huang, M.-L. C. Wu, S. R. Long, S. S. Shen, W. Qu, P. Gloersen, and K. L. Fan, "A confidence limit for the empirical mode decomposition and Hilbert spectral analysis," *Proceedings of the Royal Society of London. Series A: Mathematical, Physical and Engineering Sciences*, vol. 459, no. 2037, pp. 2317–2345, 2003.

[43] G. Rilling, P. Flandrin, P. Goncalves, and others, "On empirical mode decomposition and its algorithms," in *IEEE-EURASIP Workshop on Nonlinear Signal and Image Processing NSIP*, 2003, vol. 3, pp. 8–11.

[44] T. Gandhi, B. K. Panigrahi, and S. Anand, "A comparative study of wavelet families for EEG signal classification," *Neurocomputing*, vol. 74, no. 17, pp. 3051–3057, 2011.

[45] A. Delorme, S. Makeig, and T. Sejnowski, "Automatic artifact rejection for EEG data using high-order statistics and independent component analysis," in *International workshop on ICA (San Diego, CA)*, 2001.

[46] N. Mammone and F. C. Morabito, "Enhanced automatic artifact detection based on independent component analysis and Renyi's entropy," *Neural networks*, vol. 21, no. 7, pp. 1029–1040, 2008.

[47] N. Yeung, R. Bogacz, C. B. Holroyd, and J. D. Cohen, "Detection of synchronized oscillations in the electroencephalogram: an evaluation of methods," *Psychophysiology*, vol. 41, no. 6, pp. 822–832, 2004.

[48] A. Delorme, T. Sejnowski, and S. Makeig, "Enhanced detection of artifacts in EEG data using higher-order statistics and independent component analysis," *Neuroimage*, vol. 34, no. 4, pp. 1443–1449, 2007.

[49] X. Chen, A. Liu, H. Peng, and R. K. Ward, "A Preliminary Study of Muscular Artifact Cancellation in Single-Channel EEG," *Sensors*, vol. 14, no. 10, pp. 18370–18389, 2014.

[50] C. M. Michel and M. M. Murray, "Towards the utilization of EEG as a brain imaging tool," *Neuroimage*, vol. 61, no. 2, pp. 371–385, 2012.

[51] K. Hoedlmoser, J. Birklbauer, S. Rigler, E. Mueller, and M. Schabus, "EEG recorded during gross-motor behavior," *Brain Products Press Release*, vol. 40, pp. 11–12, 2011.

# Supplementary Material

Valentina Bono, Saptarshi Das, Wasifa Jamal, and Koushik Maharatna

<div style="text-align:center">

TABLE I
% OF THE MOST REJECTED IMF ACROSS CHANNELS

</div>

| Artifact type | IMF rejected | % across trials and channels |
|---|---|---|
| Eyeblink | 4 | 49.35 |
| Hello | 4 | 43.75 |
| Chewing | 4 | 54.12 |
| Left hand | 4 | 44.64 |
| Right hand | 4 | 40.11 |
| Left right head | 4 | 54.39 |
| Down up head | 3 | 49.29 |
| Shaking head | 3 | 51.78 |

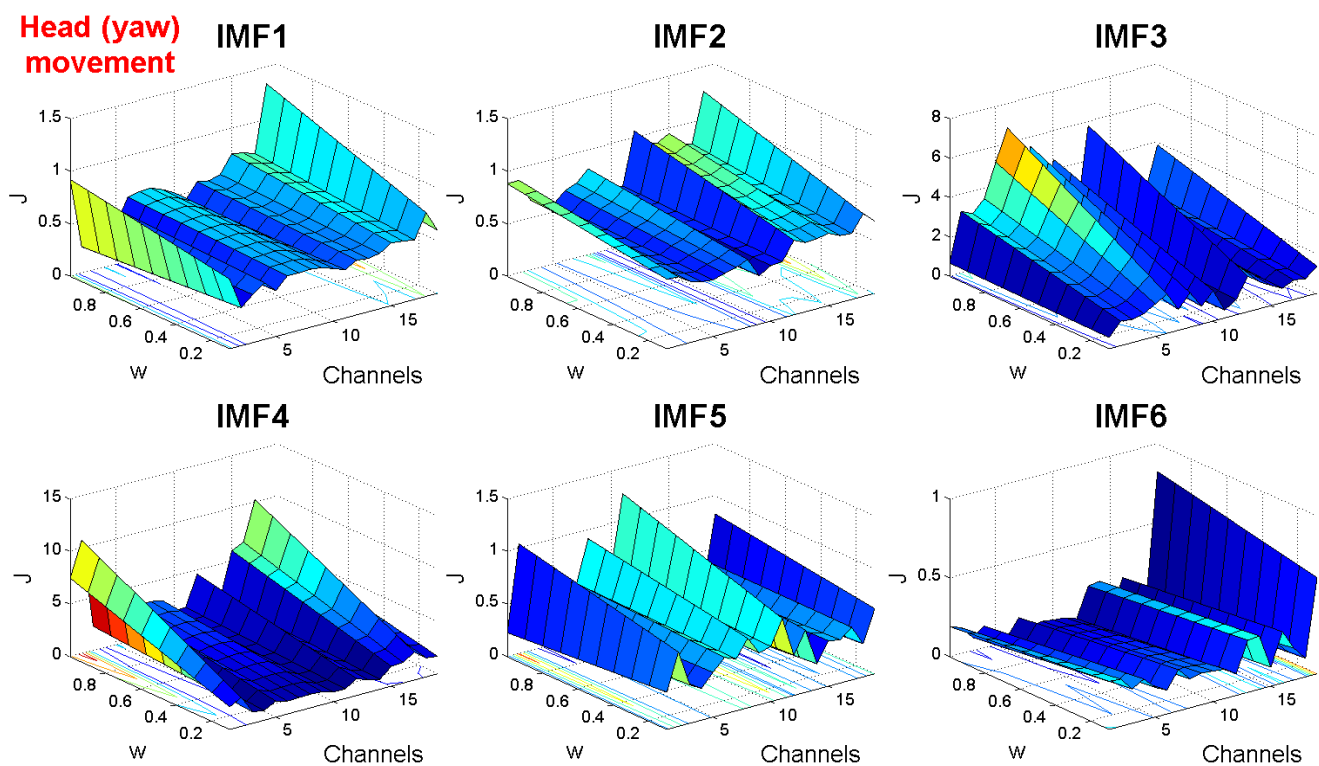

Figure 1: Surface plot of the values of the $J$ parameter across channels and IMF estimated by varying the weight $w$ for the left right (yaw) head movement artifact.

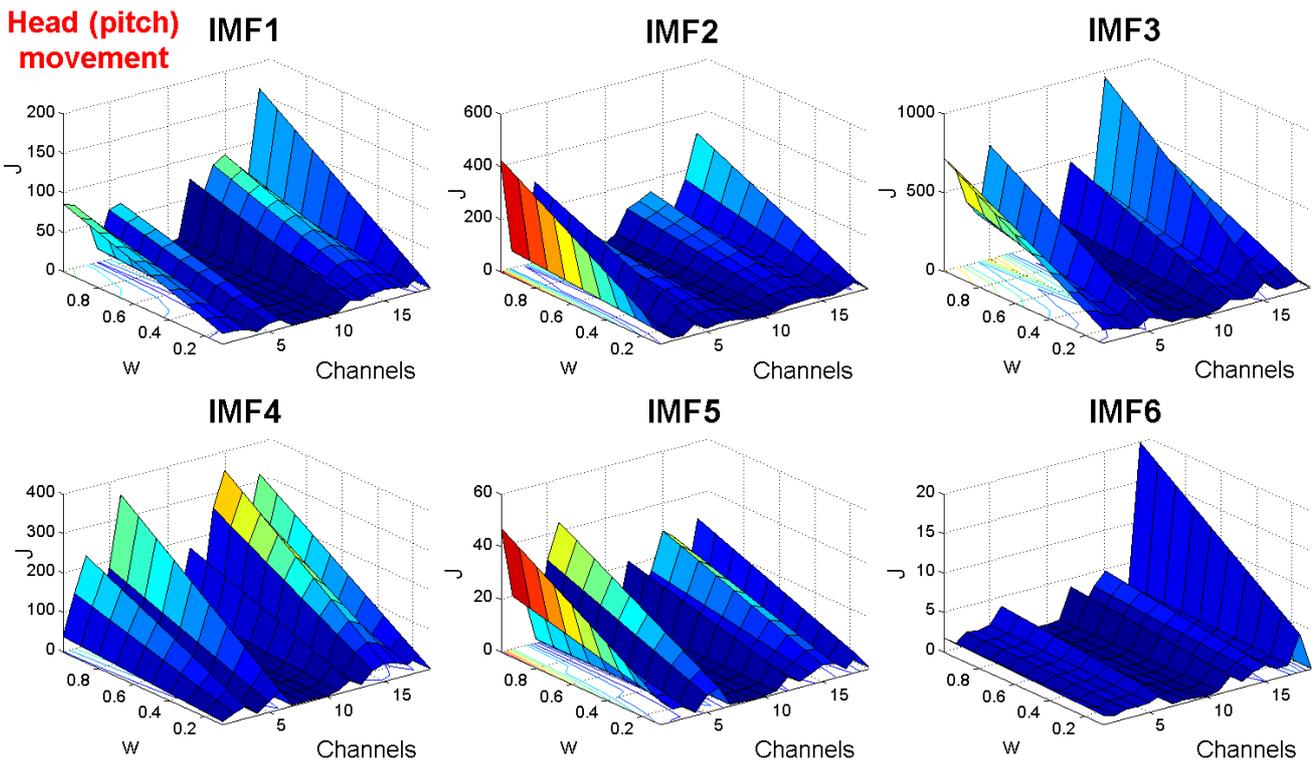

Figure 2: Surface plot of the values of the $J$ parameter across channels and IMF estimated by varying the weight $w$ for the down up (pitch) head movement artifact.

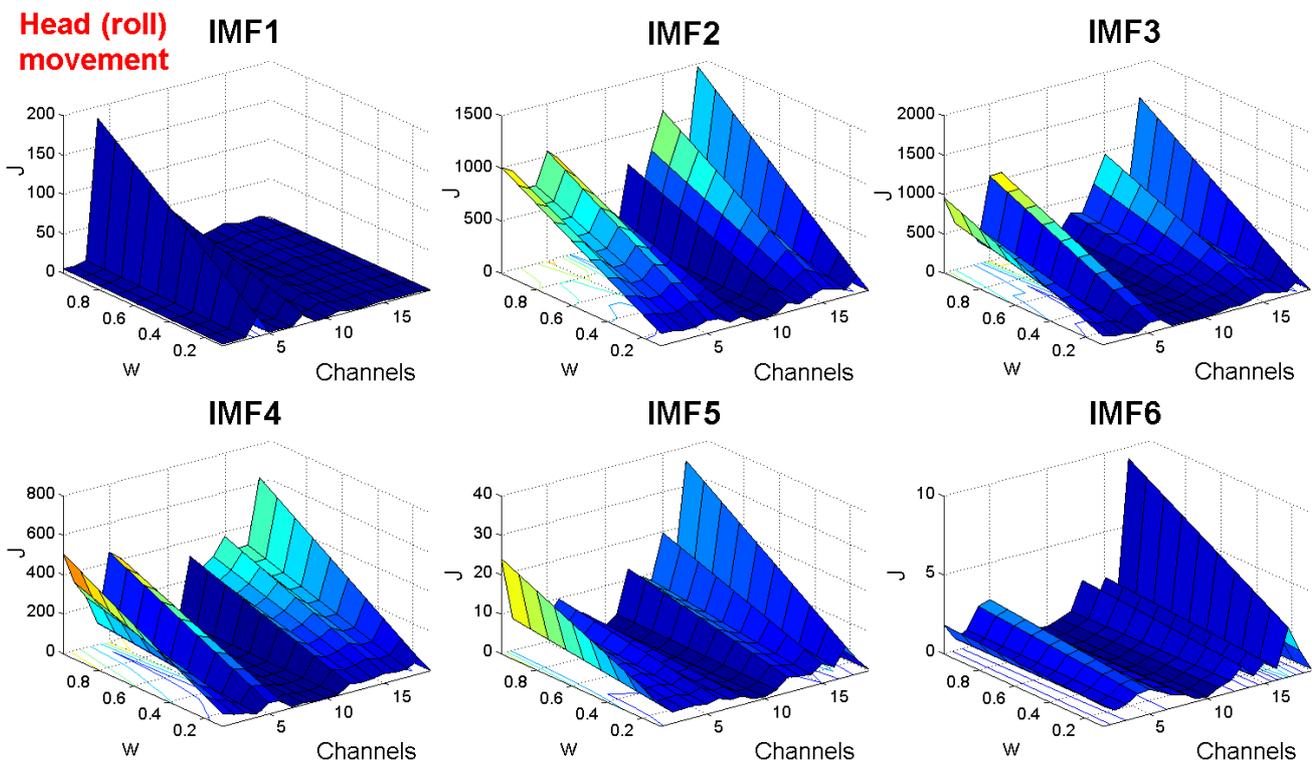

Figure 3: Surface plot of the values of the $J$ parameter across channels and IMF estimated by varying the weight $w$ for the shaking head (roll) movement artifact.

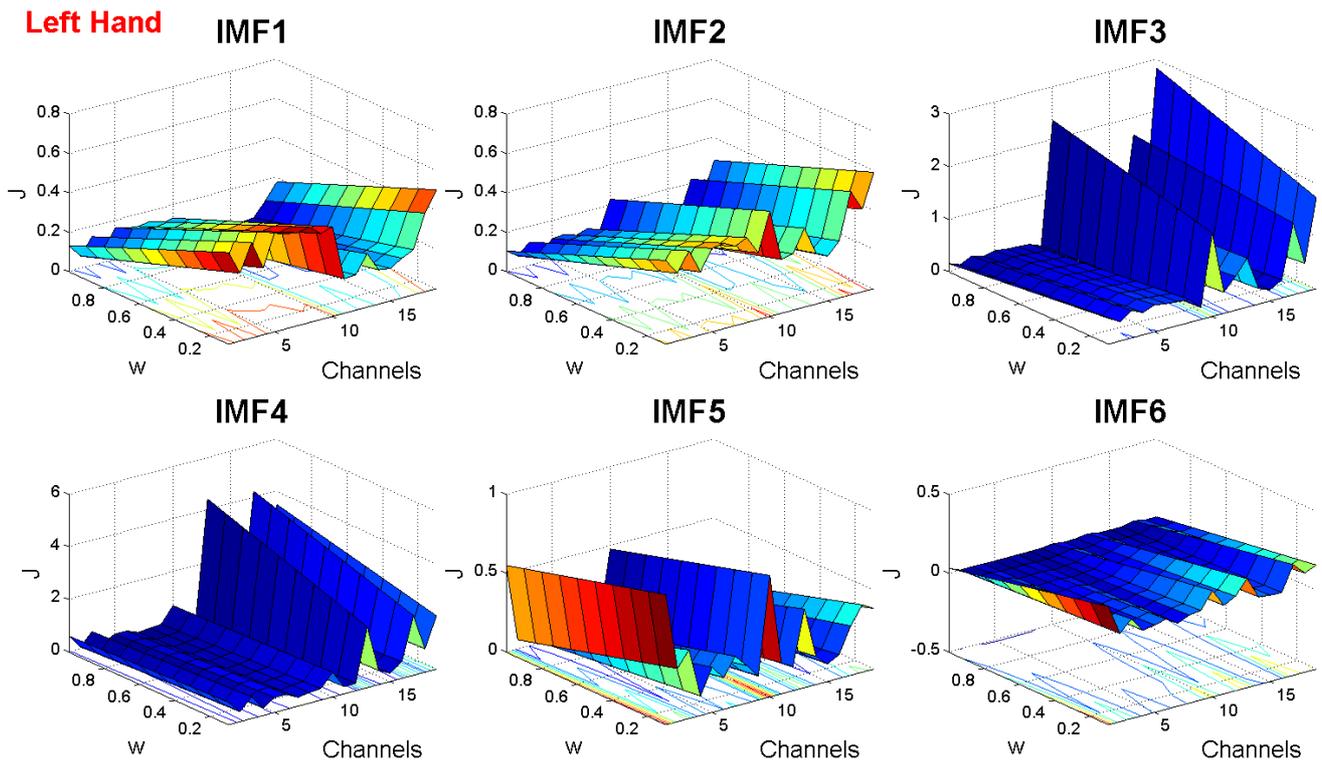

Figure 4: Surface plot of the values of the *J* parameter across channels and IMF estimated by varying the weight *w* for the left hand movement artifact.

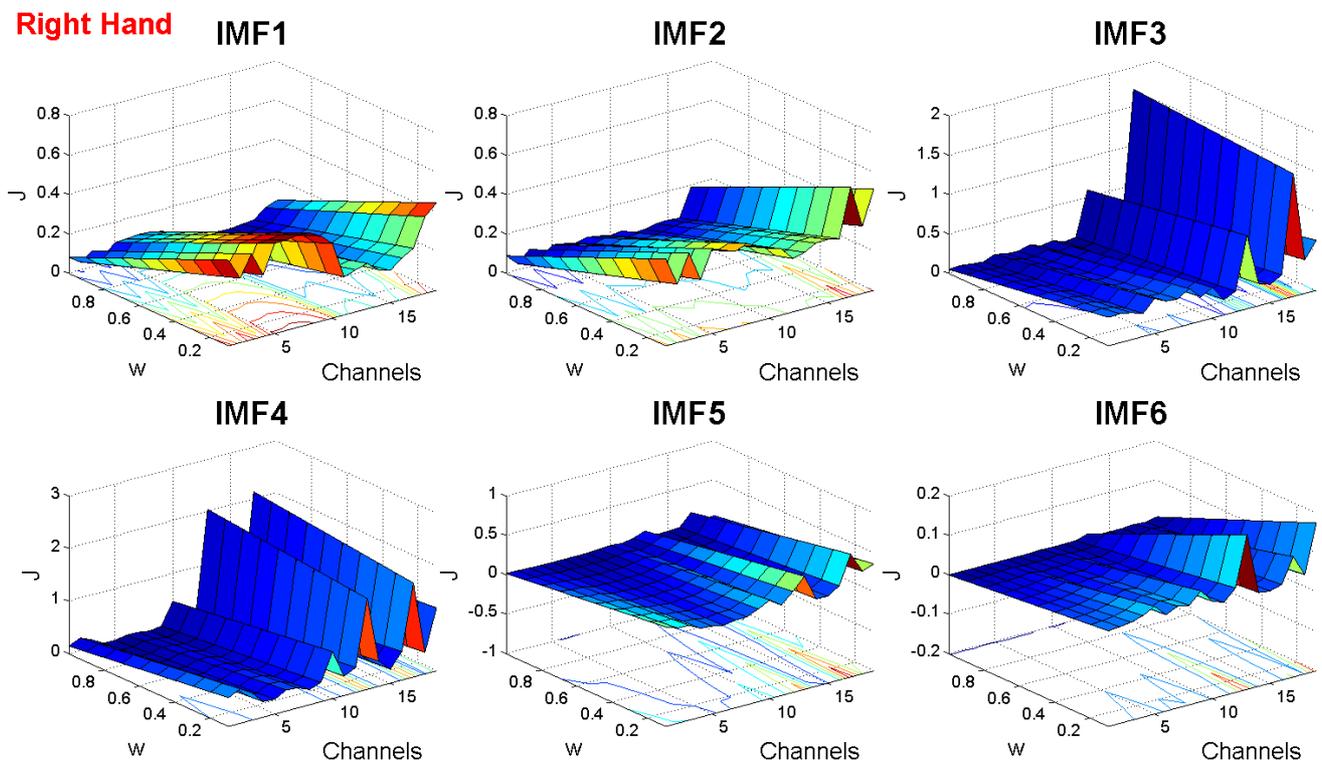

Figure 5: Surface plot of the values of the *J* parameter across channels and IMF estimated by varying the weight *w* for the right hand movement artifact.

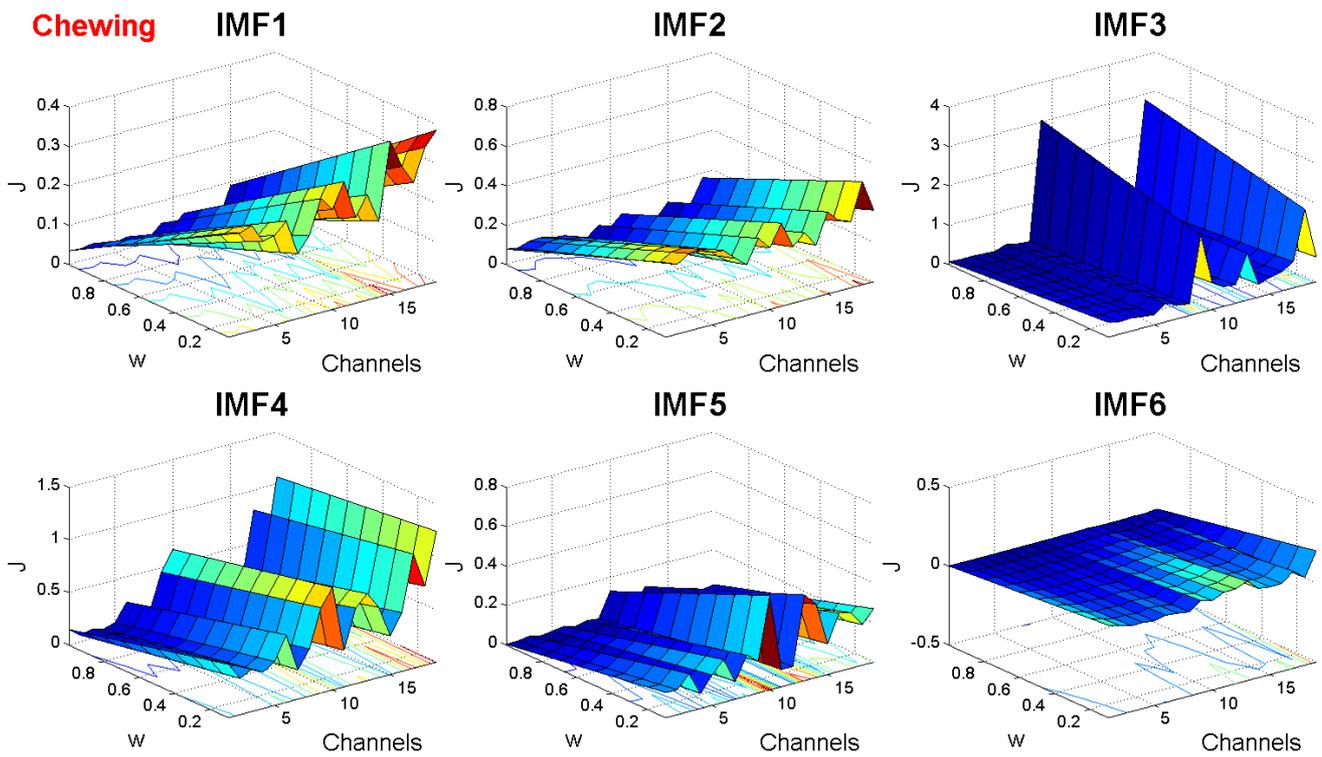

Figure 6: Surface plot of the values of the $J$ parameter across channels and IMF estimated by varying the weight $w$ for the chewing artifact.

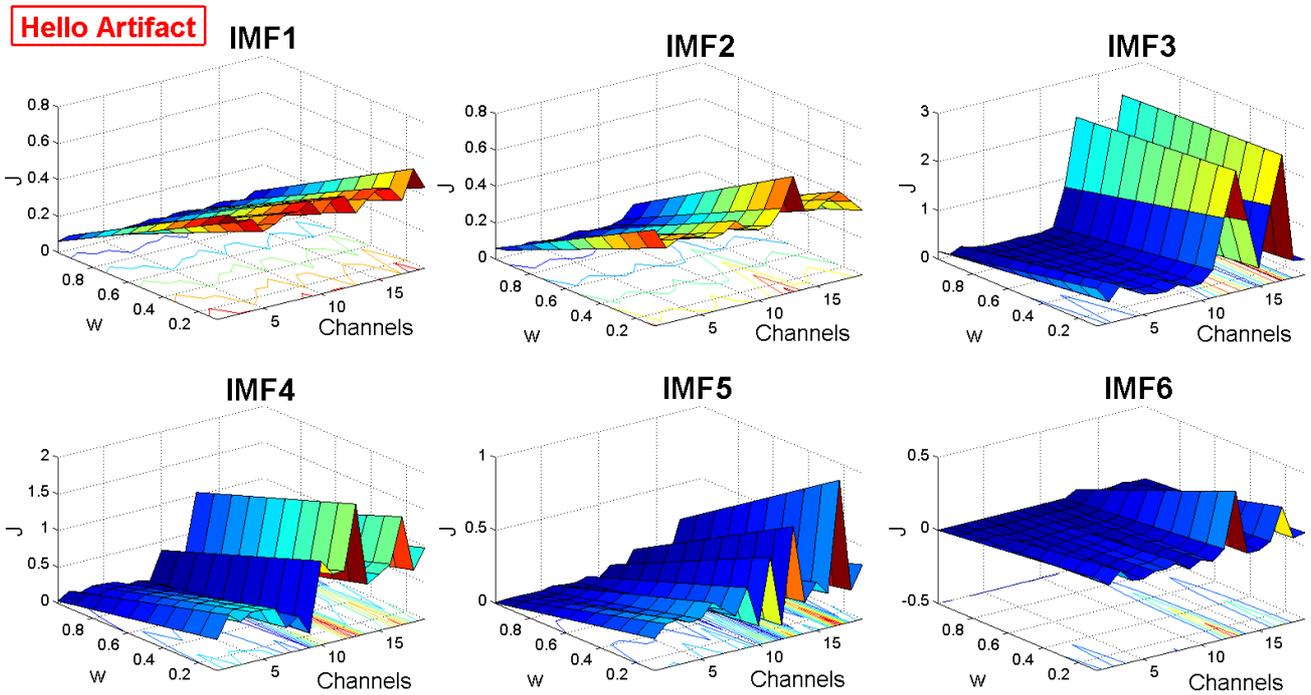

Figure 7: Surface plot of the values of the $J$ parameter across channels and IMF estimated by varying the weight $w$ for the real speaking (saying "hello") artifact.

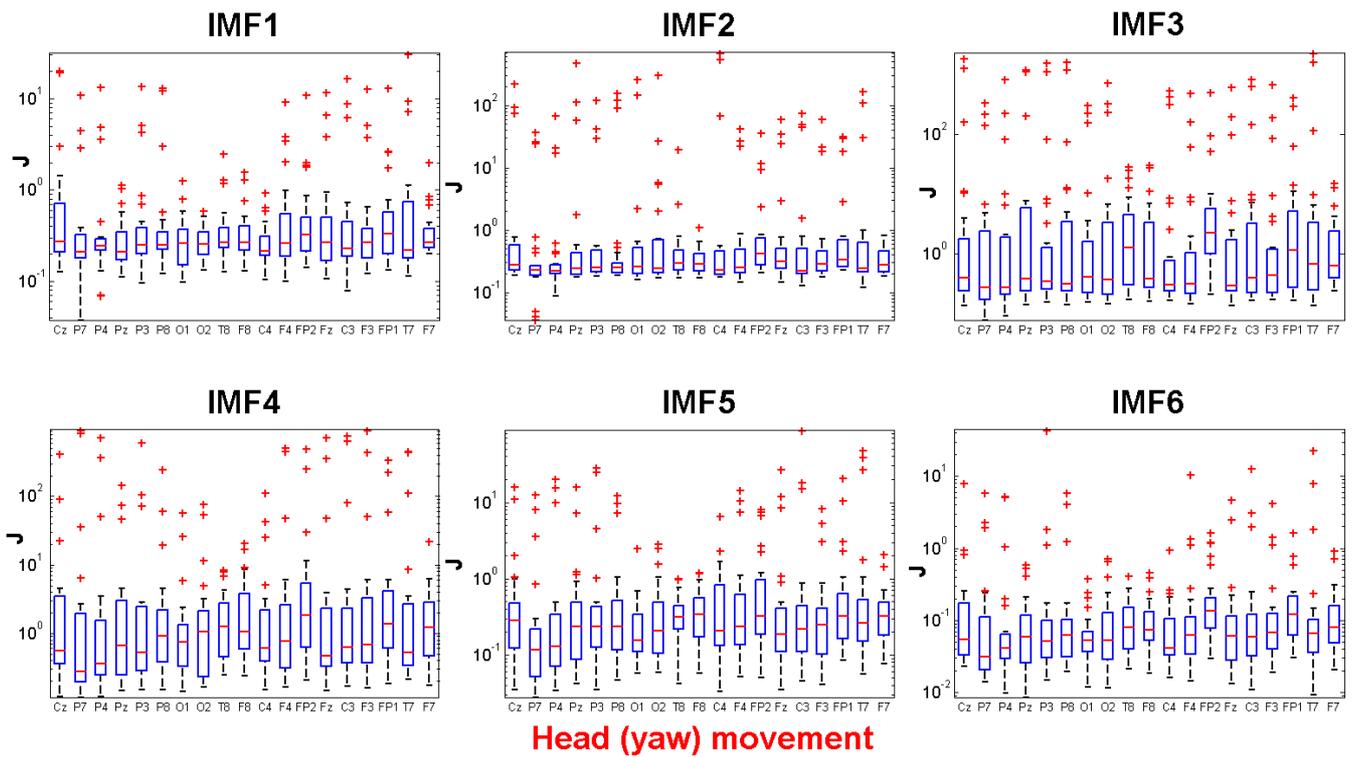

Figure 8: Example of IMF rejected across all trials for the left right (yaw) head movement artifact.

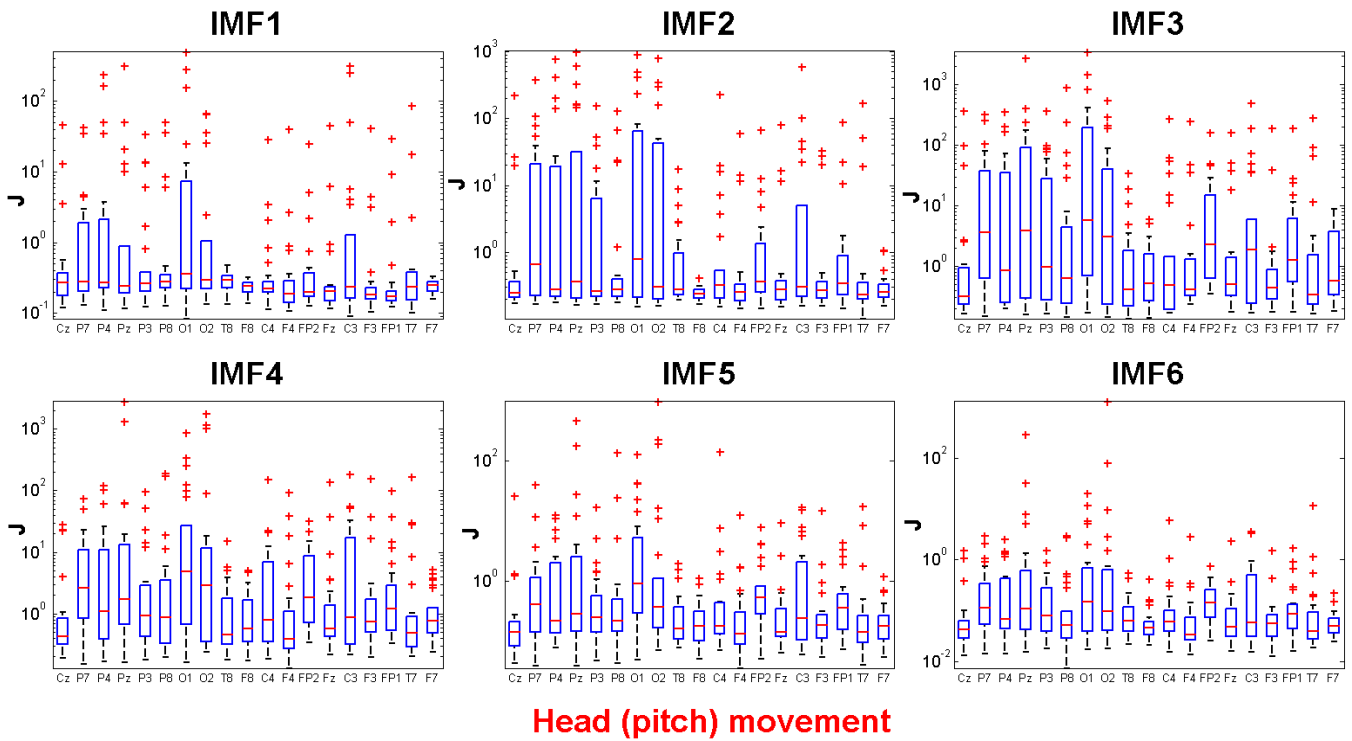

Figure 9: Example of IMF rejected across all trials for the down up (pitch) head movement artifact.

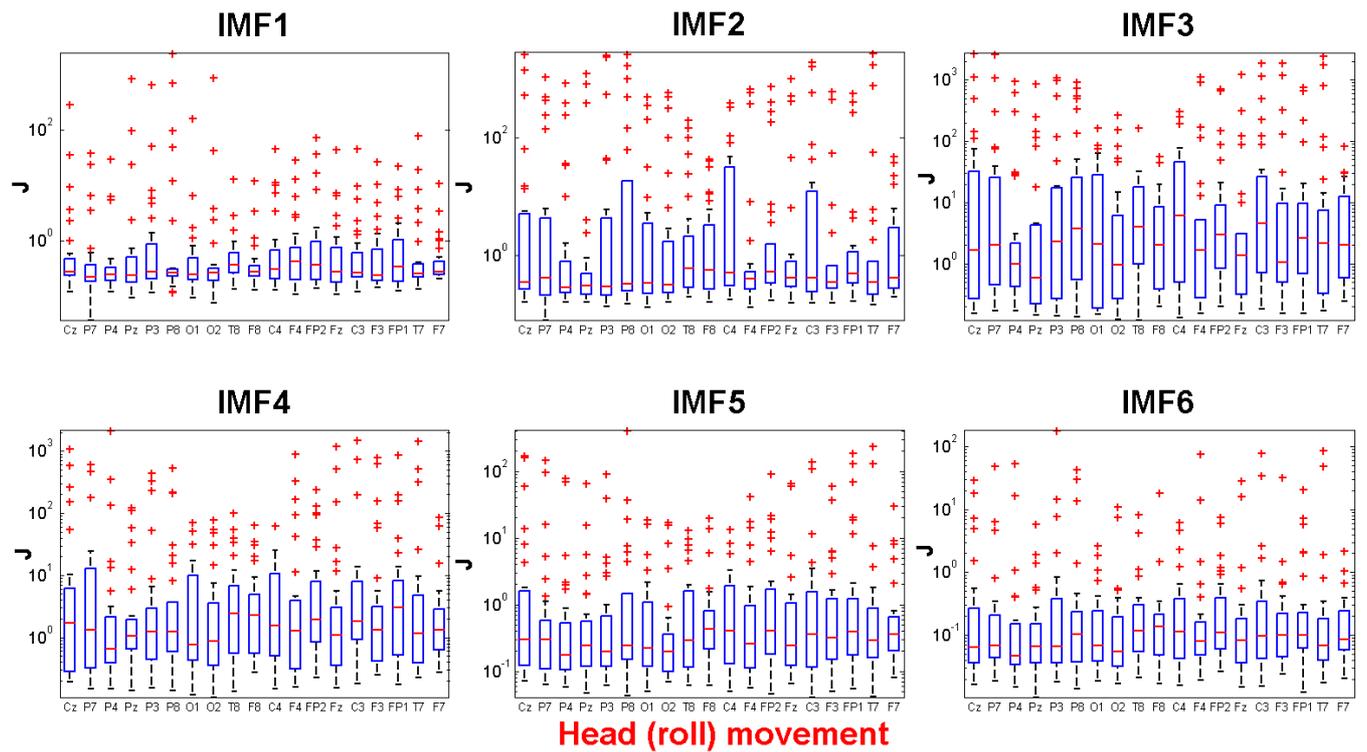

Figure 10: Example of IMF rejected across all trials for the shaking head (roll) movement artifact.

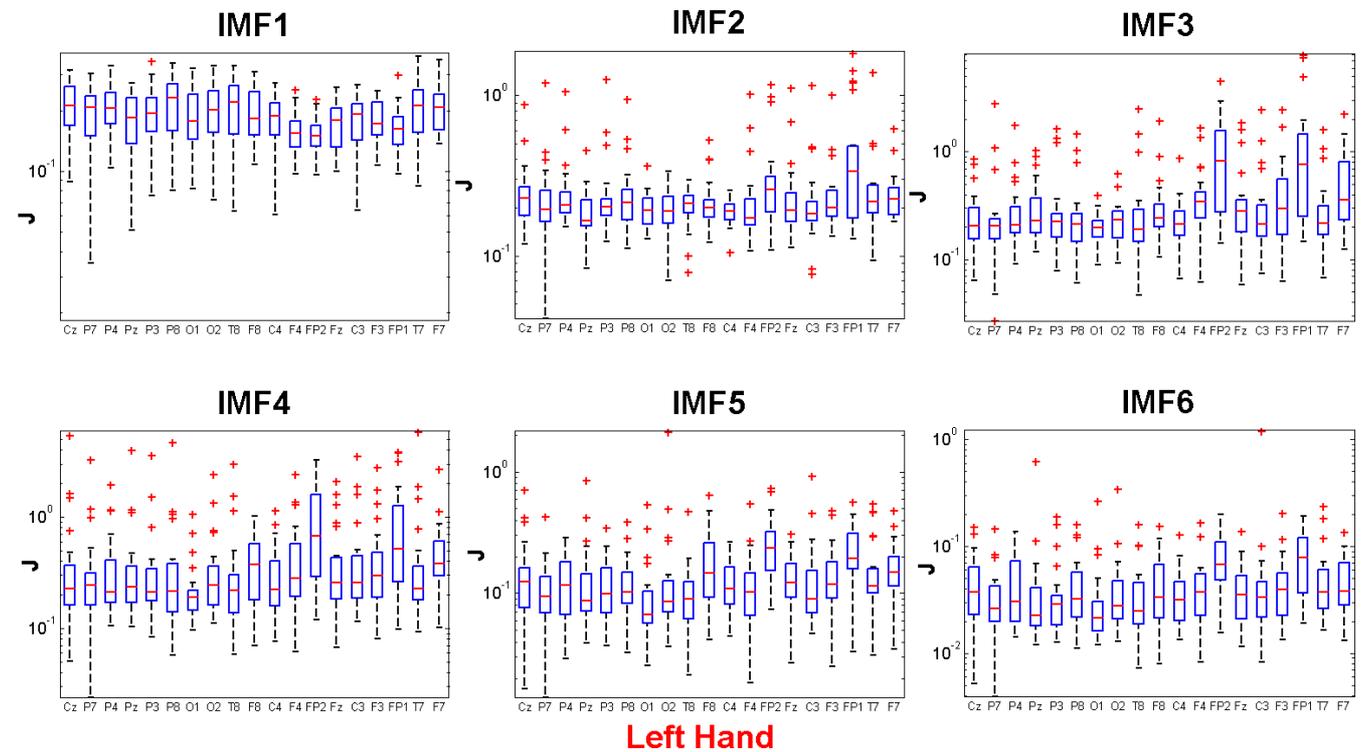

Figure 11: Example of IMF rejected across all trials for the left hand movement artifact.

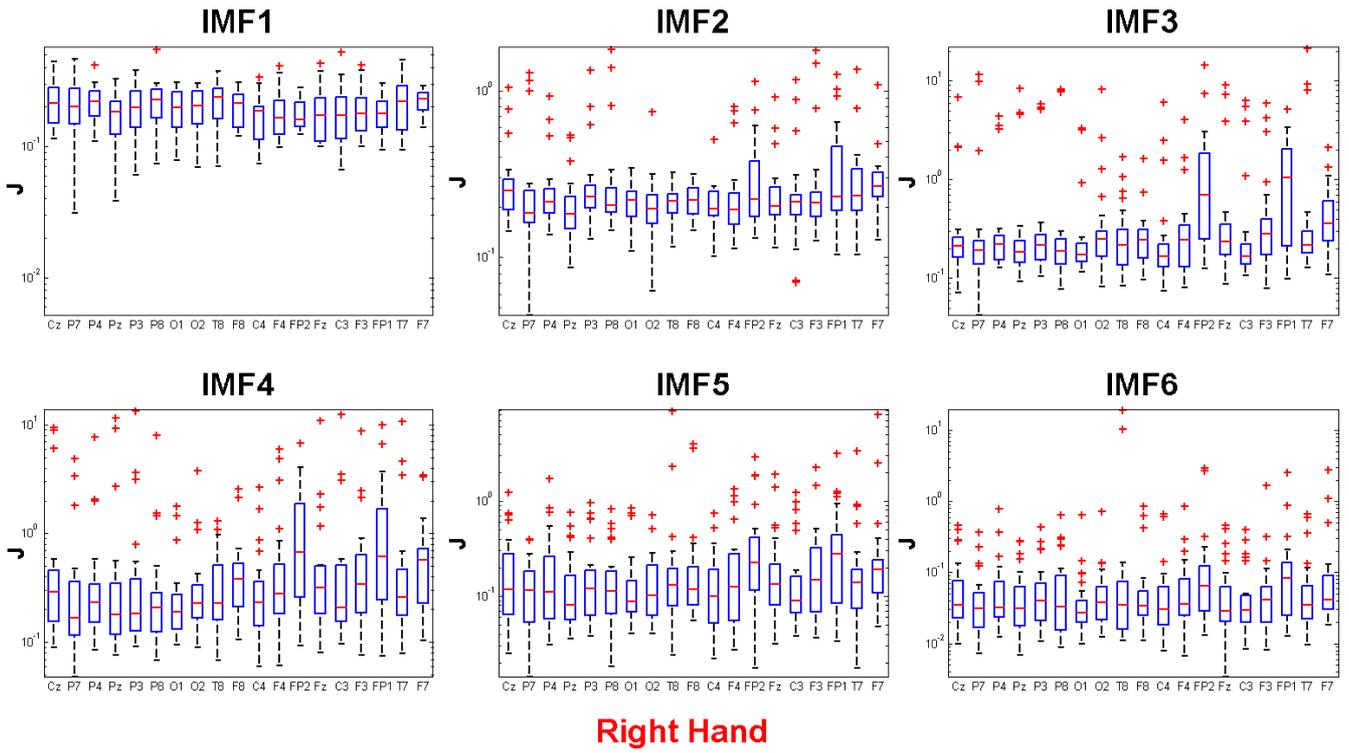

**Right Hand**

Figure 12: Example of IMF rejected across all trials for the right hand movement artifact.

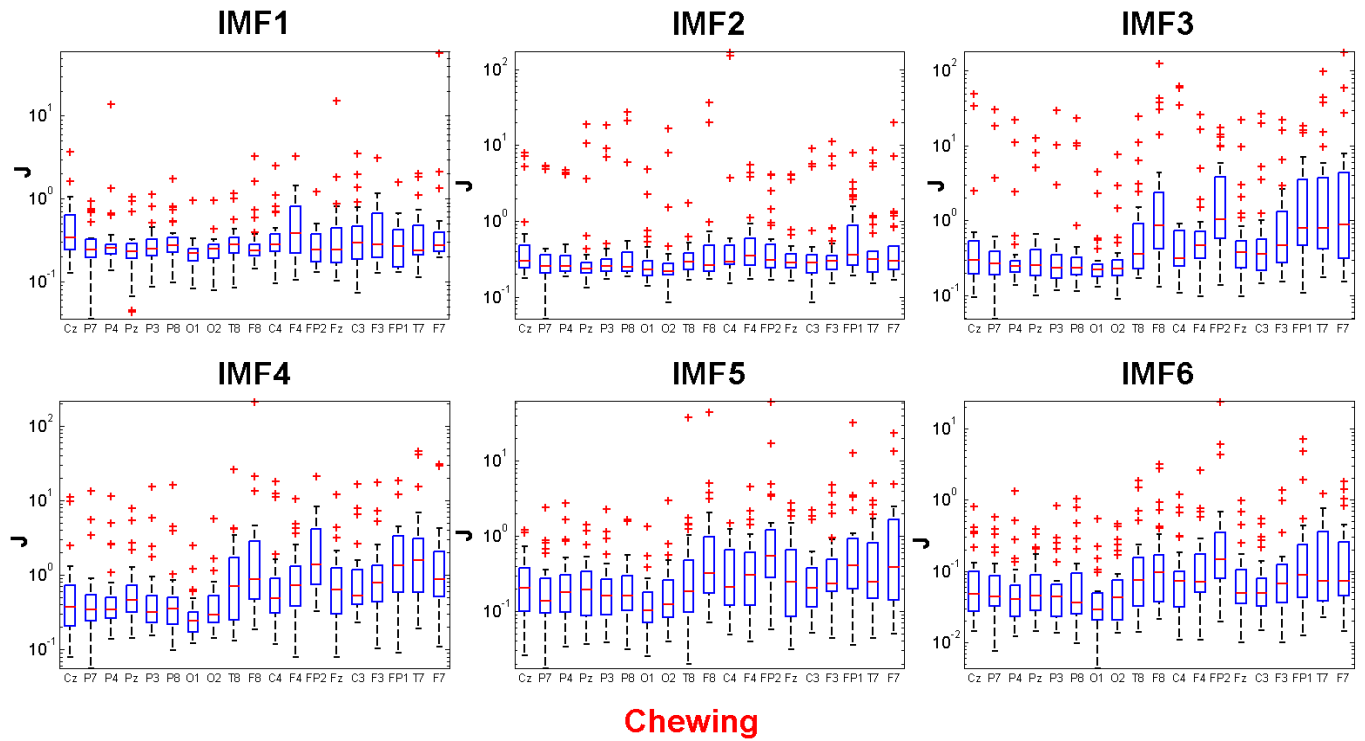

**Chewing**

Figure 13: Example of IMF rejected across all trials for the chewing artifact.

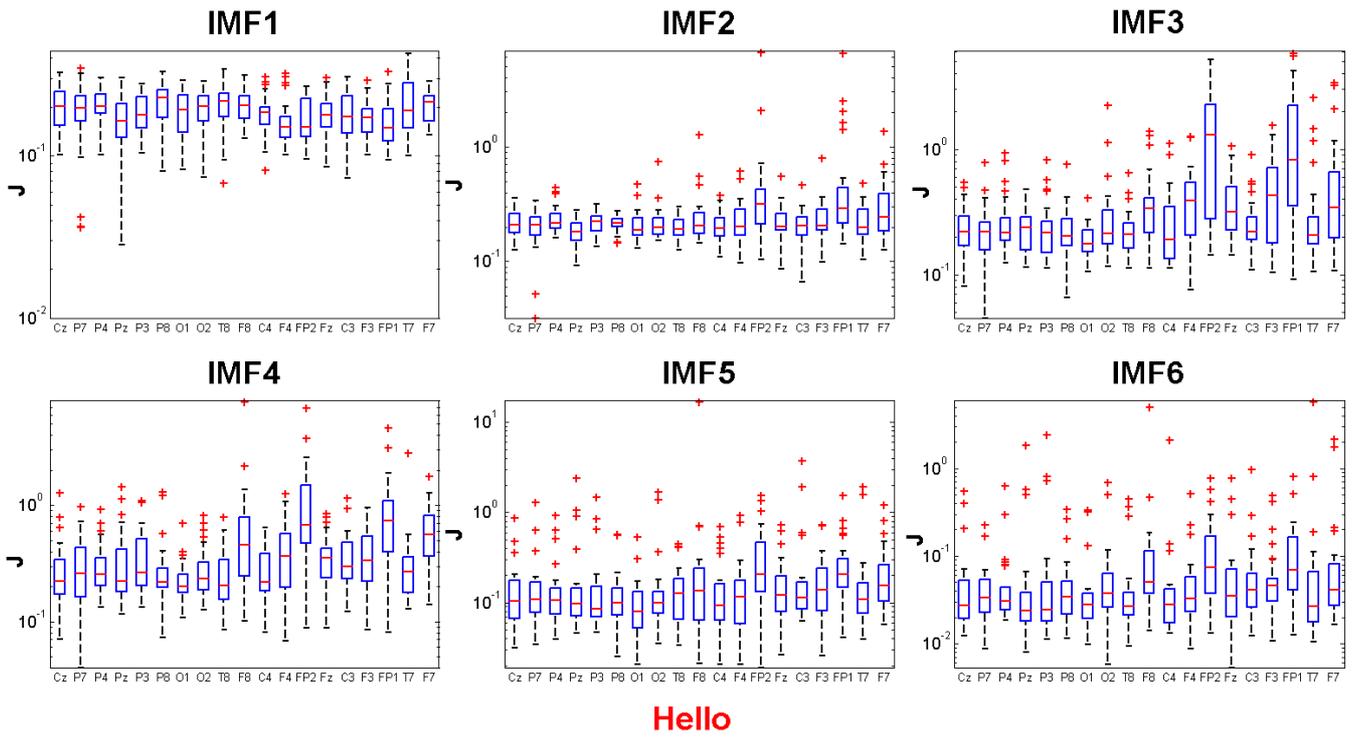

Figure 14: Example of IMF rejected across all trials for the hello artifact.

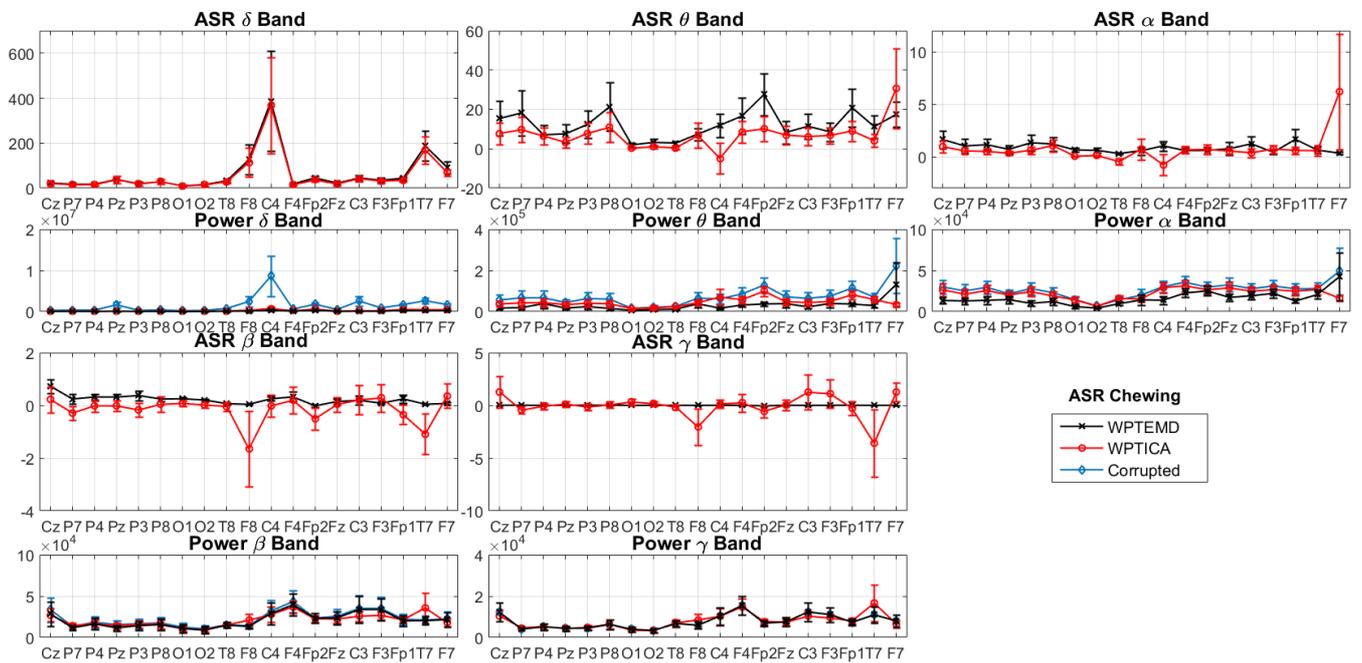

Figure 15: ASR and power for all trials of real chewing across all the subjects for each band.

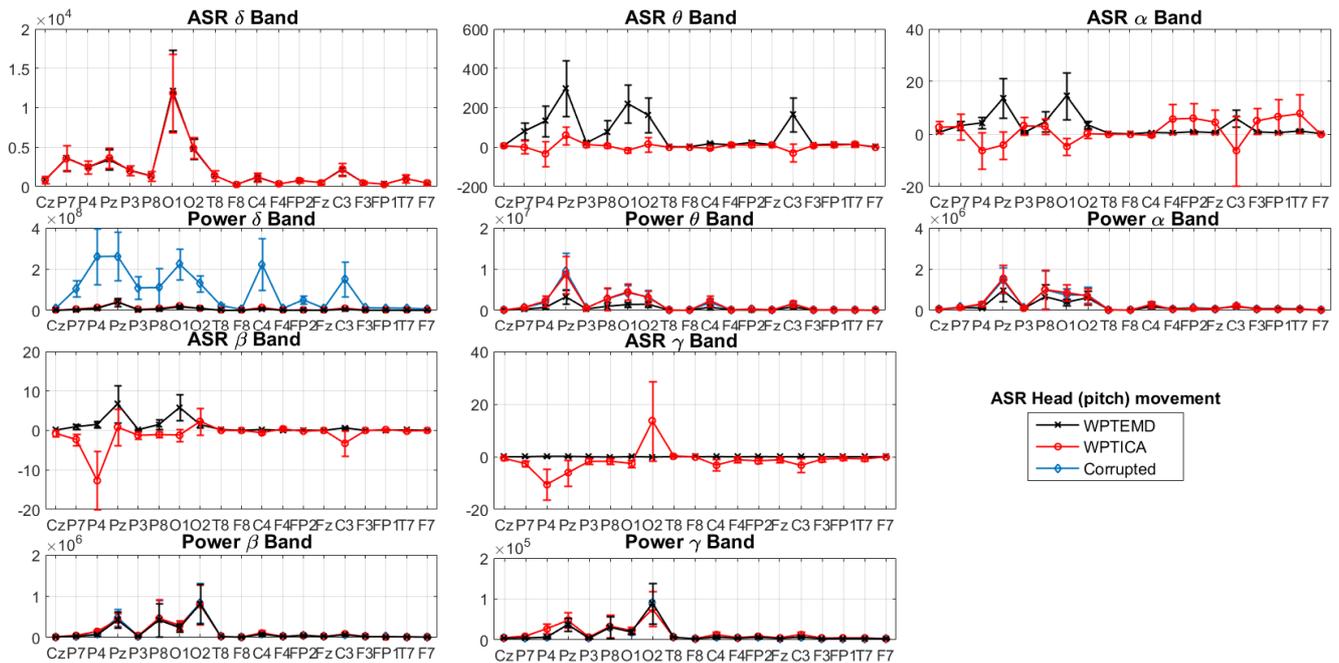

Figure 16: *ASR* and power for all trials of real head (pitch) movement across all the subjects for each band.

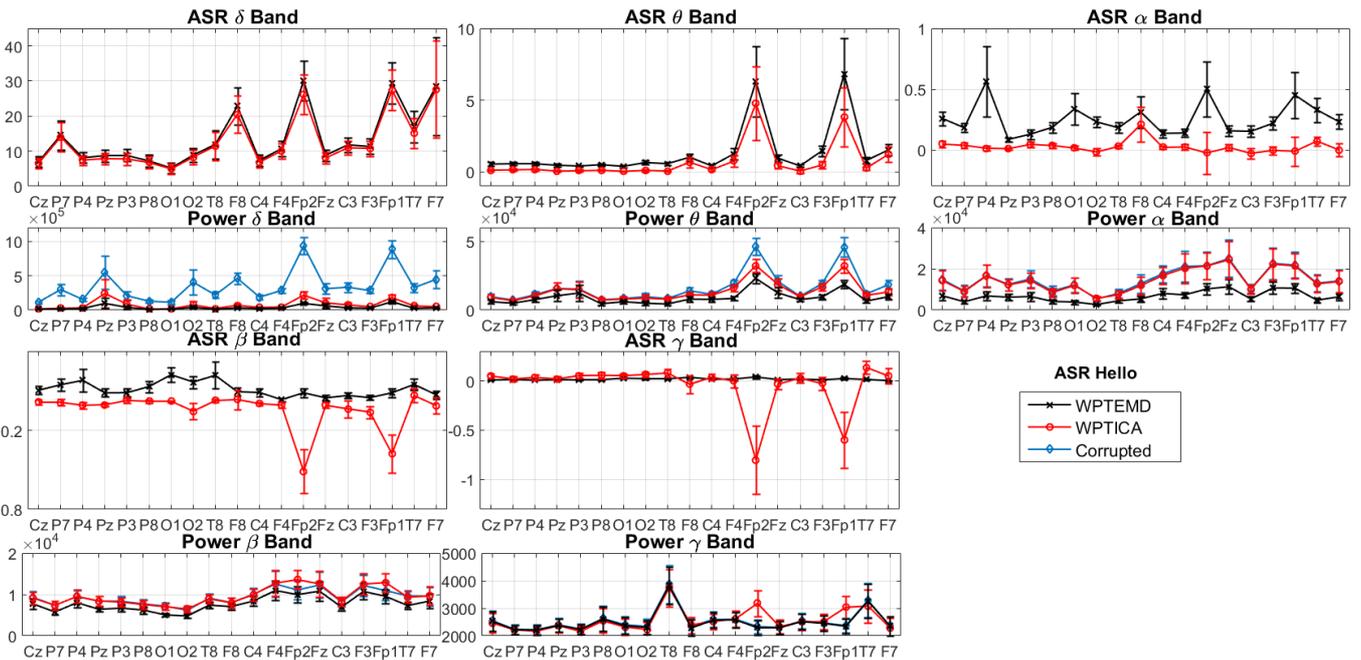

Figure 17: *ASR* and power for all trials of real speaking (i.e. saying "hello") across all the subjects for each band.

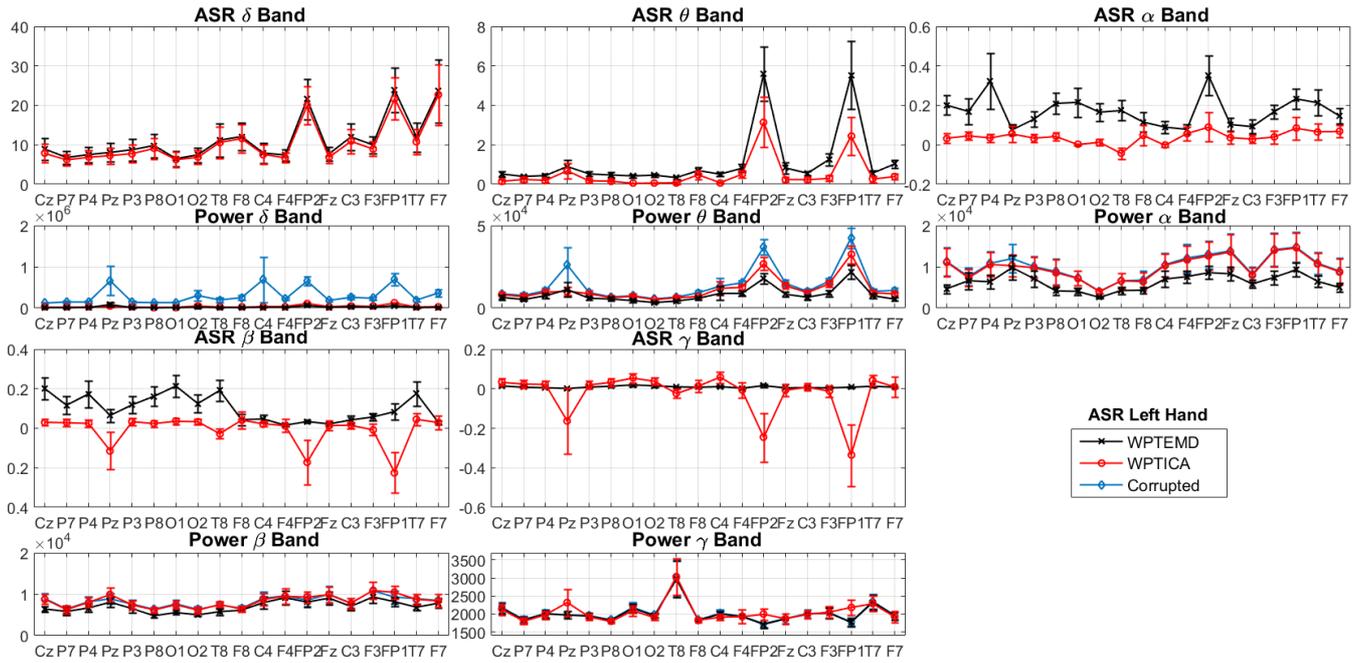

Figure 18: *ASR* and power for all trials of real left hand movement across all the subjects for each band.

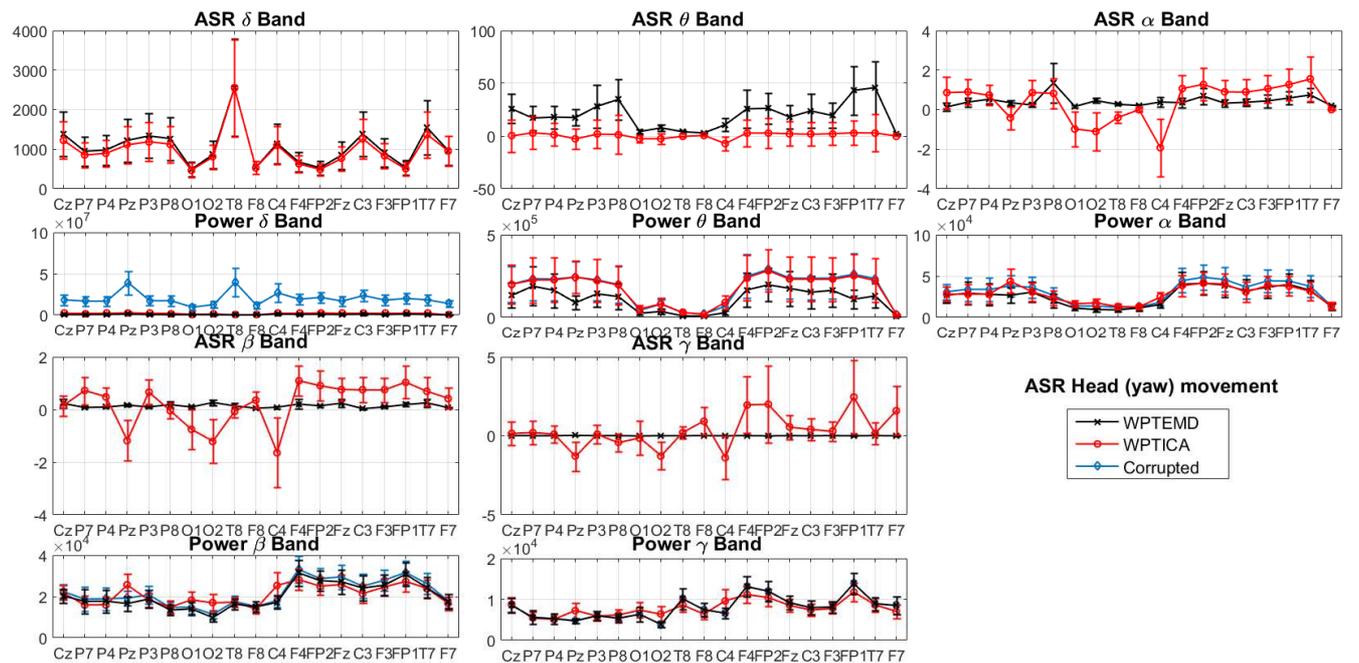

Figure 19: *ASR* and power for all trials of real head (yaw) movement across all the subjects for each band.

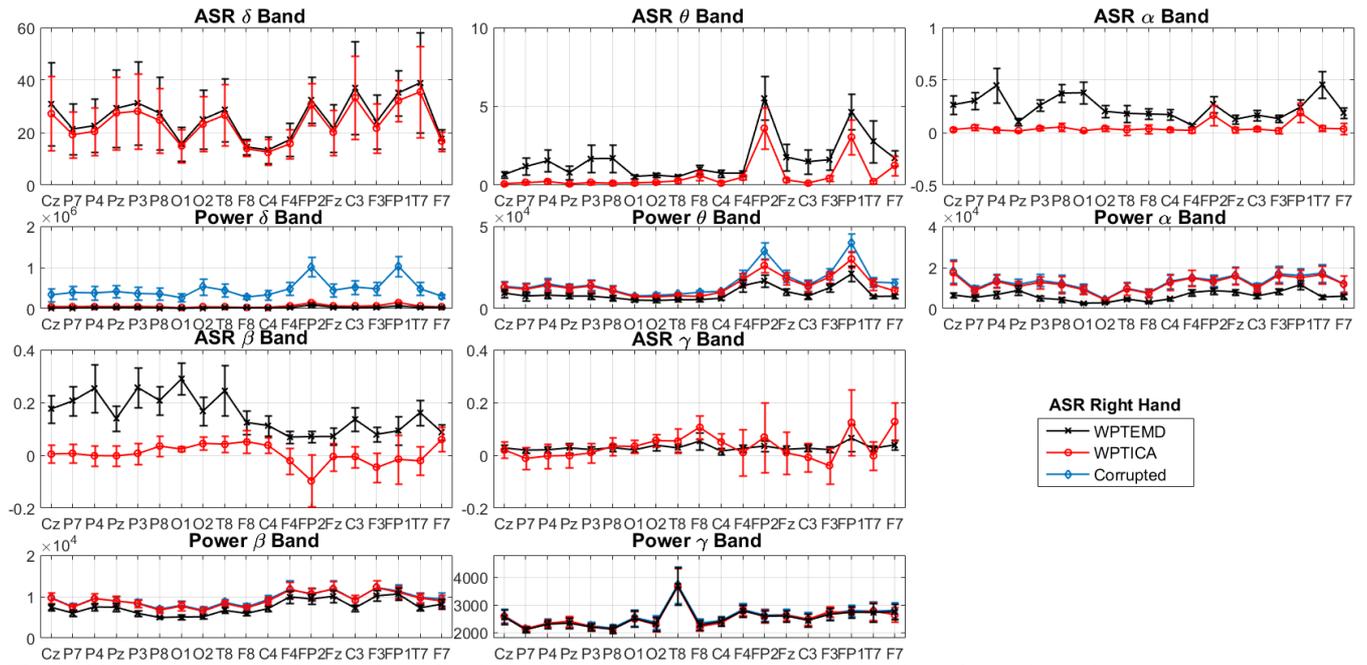

Figure 20: *ASR* and power for all trials of real right hand movement across all the subjects for each band.

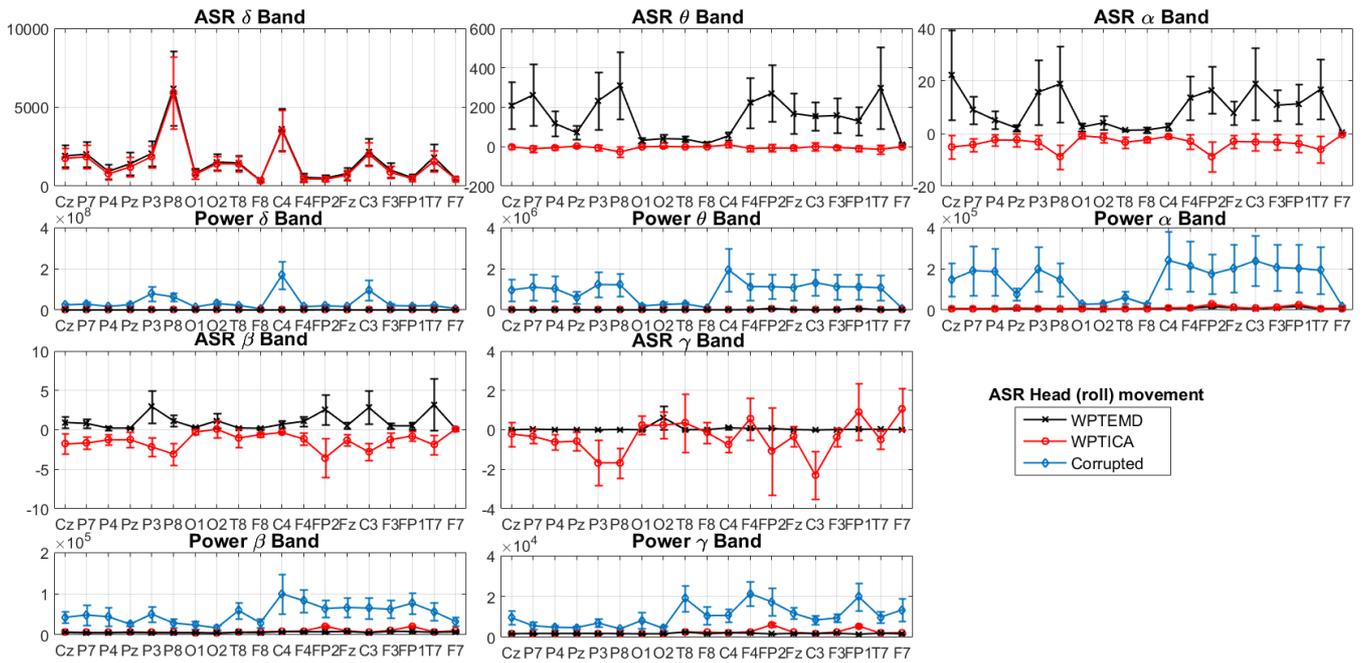

Figure 21: *ASR* and power for all trials of real head (roll) movement across all the subjects for each band.

# Supplementary Material on Single Subject Single Trial EEG Based Scalp Topography Analysis

Valentina Bono, Saptarshi Das, Wasifa Jamal, Koushik Maharatna

Figure 1 shows the topographical maps during eye-blink artifact. From the scalp topography across all bands (i.e. *All bands*) of the corrupted signal, it is evident that only the frontal region shows high power. The same pattern is present in all the frequency bands except $\gamma$, suggesting that the artifact affects the EEG power in the frequency range up to 30 Hz. The WPT and WPTICA algorithms are capable of reducing most of the power in the frontal electrodes only in the $\delta$ band, while the WPTEMD technique reduces the power in each frequency band except for the $\gamma$ band which is less affected by eye-blinking.

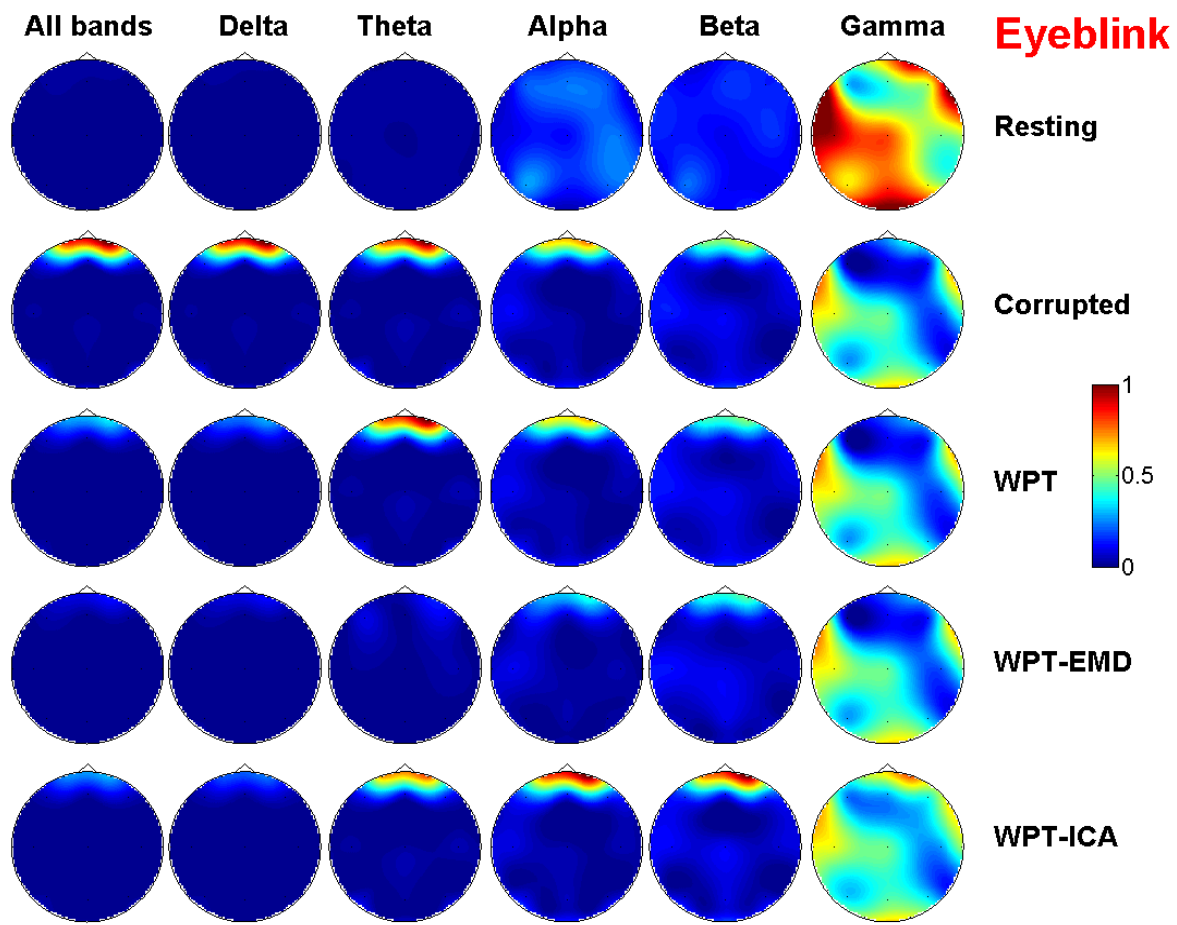

Figure 1: Distribution of the scalp topography power (various bands) for single subject-single trial of eye-blink.

The yaw movement of the head affects the left and right temporal electrodes, as visible in the scalp topography-*All bands* in Figure 2, and the same pattern is present in the $\delta$ and $\theta$ bands. It is fully removed in $\delta$ band by the WPT algorithm and partially removed in $\theta$ band by the WPTEMD technique.

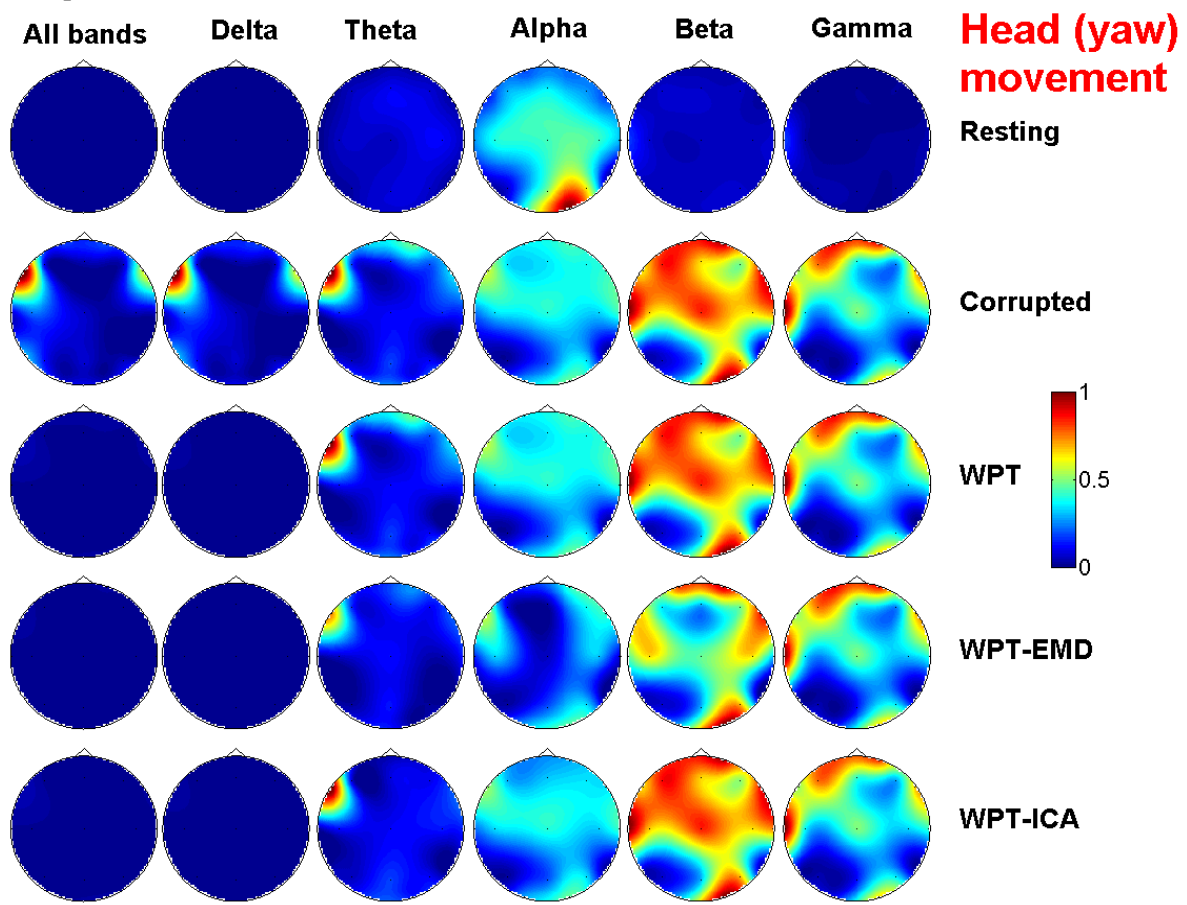

Figure 2: Distribution of the scalp topography power (various bands) for single subject-single trial of head (yaw) movement.

The head (pitch) movement generates high power in the occipital electrodes, as visible in the scalp topography-*All bands* in Figure 3. The occipital region shows high power in all the frequency bands, suggesting that the artifact affects all of them. The WPTEMD algorithm is the only technique capable of suppressing the high power in all the bands, except the $\gamma$ which shows the same pattern in the corrupted signal even after the application of the WPT and WPTEMD techniques. The power in $\beta$ and $\gamma$ bands increases when using the WPTICA and hence would not be recommended here.

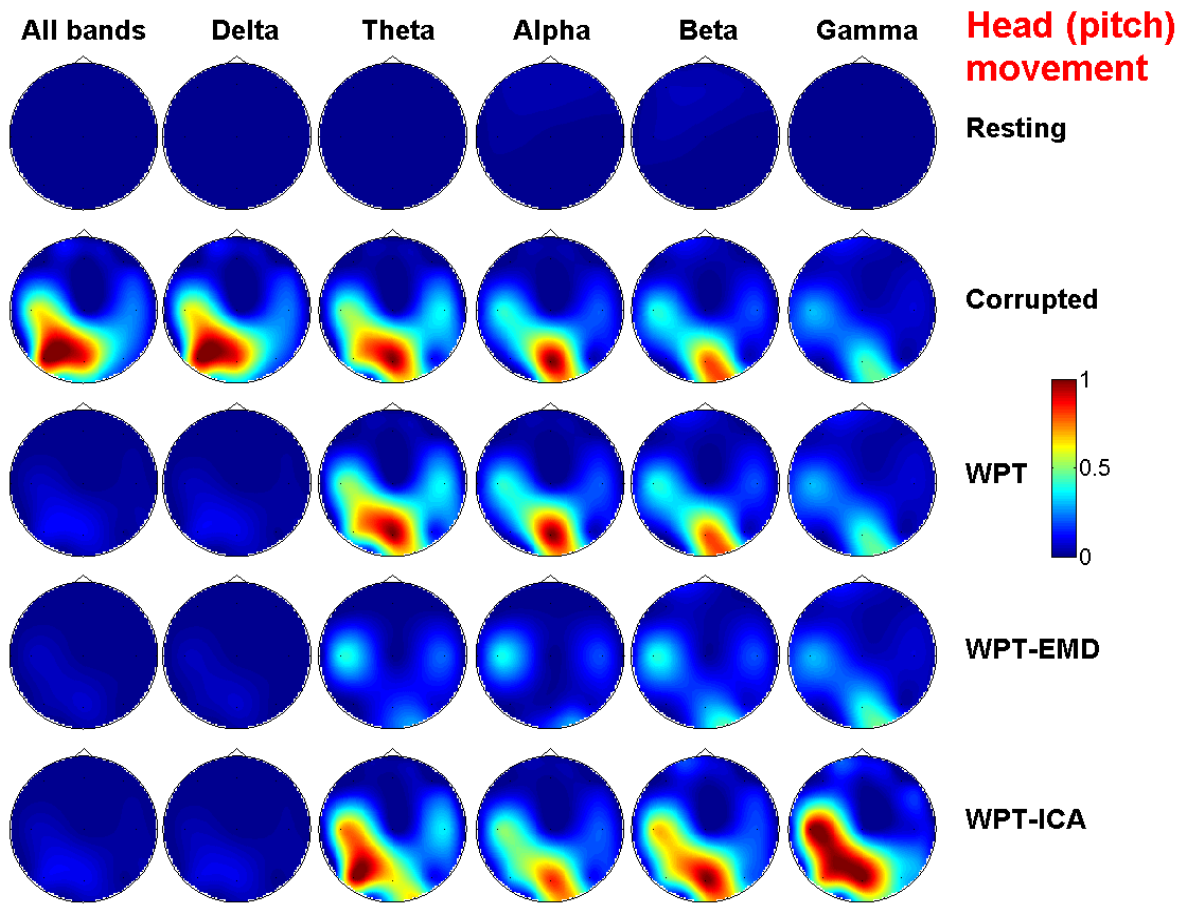

Figure 3: Distribution of the scalp topography power (various bands) for single subject-single trial of head pitch movement.

Head (roll) movement affects the temporal lobes, as shown in the scalp topography-*All bands* in Figure 4. Similar topography pattern is visible in the $\delta$ band, while the other bands show different power over the scalp. Both the WPTEMD and WPTICA algorithms are capable of removing the artifact in the $\delta$ band, due to the action of the WPT technique. The WPTEMD is also capable of reducing the power in the $\theta$ and $\alpha$ bands, whereas these bands are less modified by WPTICA.

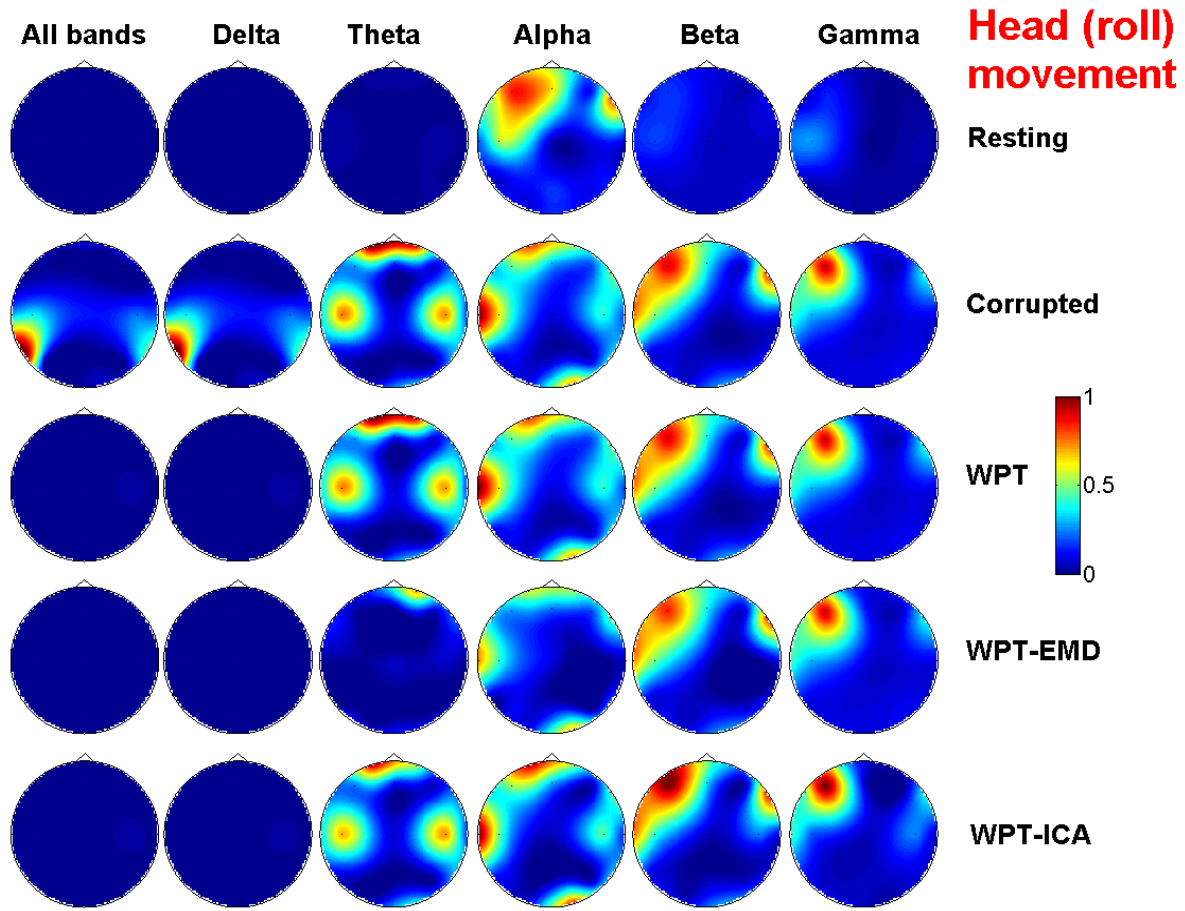

Figure 4: Distribution of the scalp topography power (various bands) for single subject-single trial of head roll movement.

Figure 5 shows that the highest power is located in the frontal and left temporal regions for the left hand movement artifact. The high power in the scalp topography-*All bands* is also visible in the $\delta$, $\theta$ and $\alpha$ bands, suggesting that the artifact might affect the power in the frequency range up to 13 Hz, which is mostly suppressed by the WPTEMD algorithm. Unlike the WPTEMD, the scalp topography in WPT and WPTICA still contains high power in the frontal region.

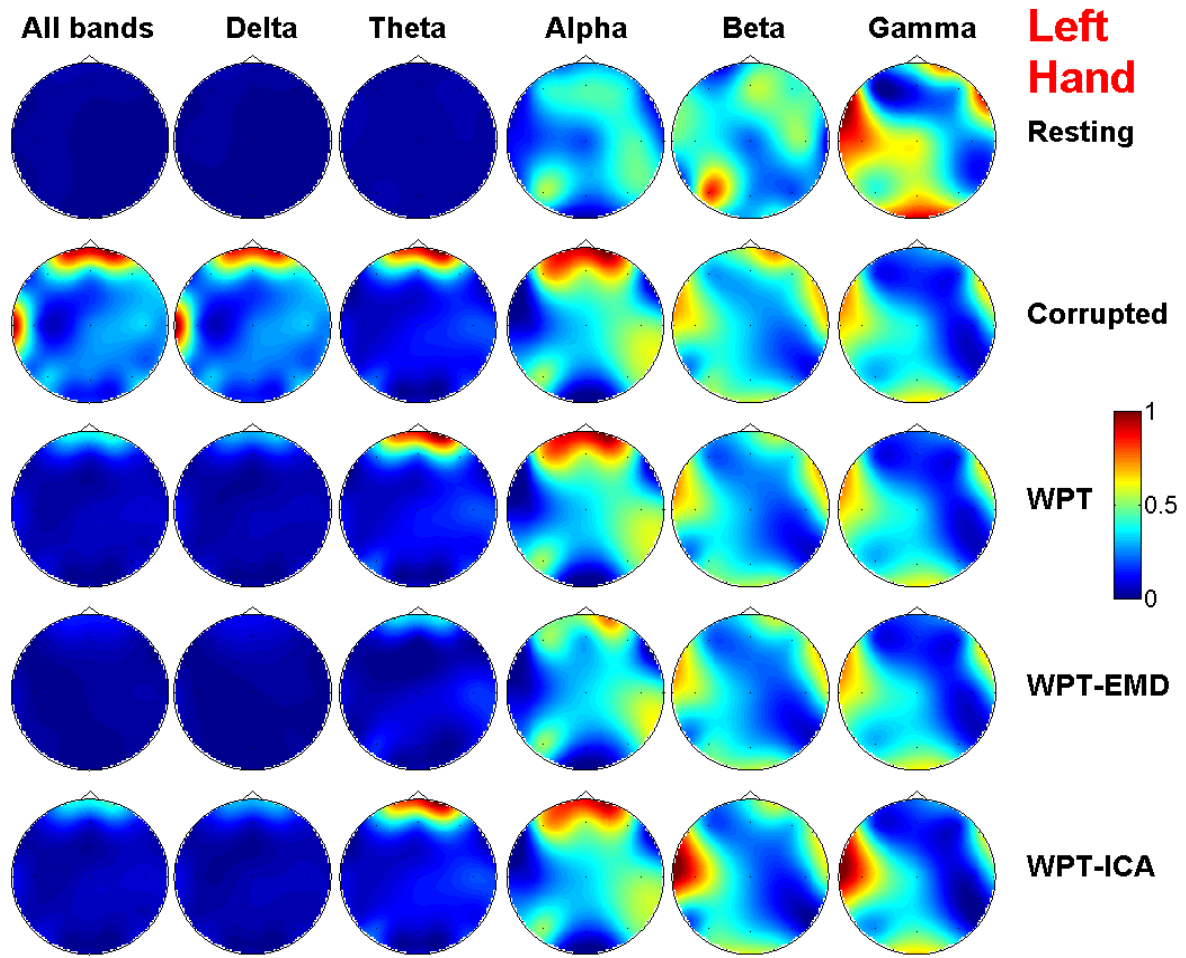

Figure 5: Distribution of the scalp topography power (various bands) for single subject-single trial of left hand movement.

The scalp topography across all bands from data recorded while the subject was performing right hand movement shows similar pattern to the previous artifact, as can be seen in Figure 6. The high power in the frontal and temporal regions shown in scalp topography is also visible in the $\delta$, $\theta$ and $\alpha$ bands and it is reduced using all the techniques. However, the $\beta$ and $\gamma$ bands are characterized by high power which is reduced only by the WPTICA algorithm.

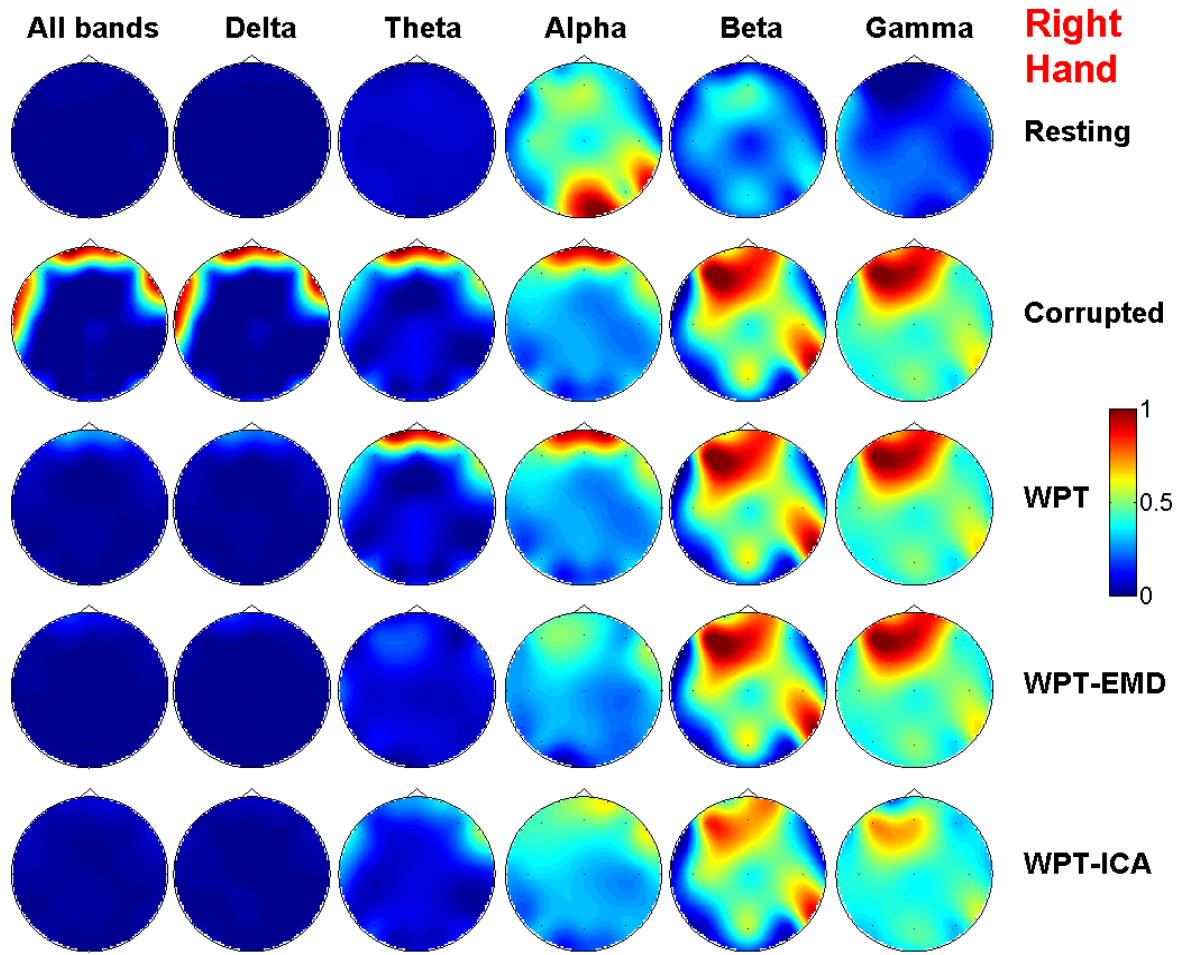

Figure 6: Distribution of the scalp topography power (various bands) for single subject-single trial of right hand movement.

Figure 7 shows that the left temporal lobe has the highest power in case of chewing, as visible in the scalp topography-*All bands*. The same pattern is detectable in the $\delta$ band and it is reduced by all the techniques. The high power characterizing the rest of the bands is reduced in a higher extent by the WPTEMD algorithm compared to the other variants.

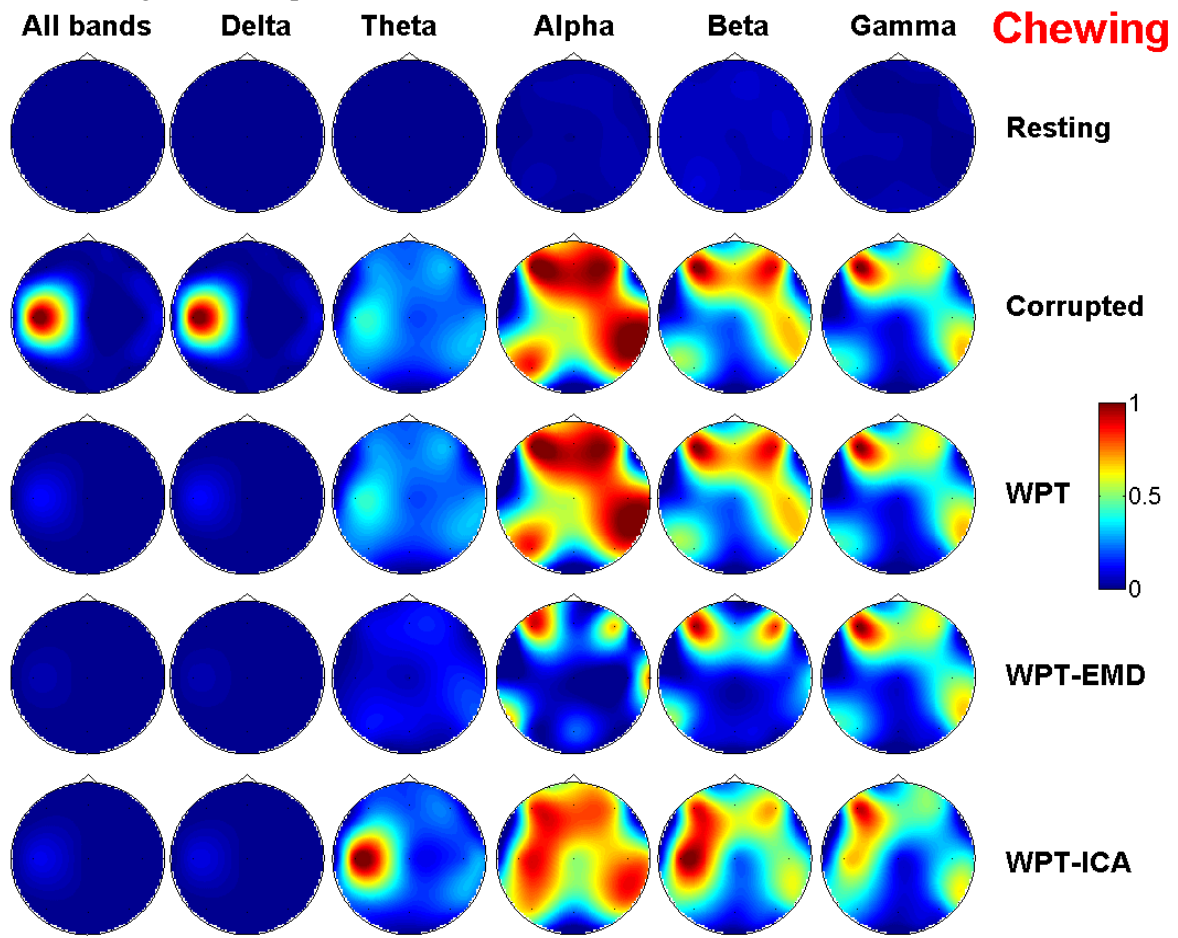

Figure 7: Distribution of the scalp topography power (various bands) for single subject-single trial of chewing.

Talking affects the frontal and left temporal regions, as shown in the scalp topography-*All bands* of Figure 8. These high power patterns in the frontal and left temporal regions are also in the $\delta$ band, in a greater extent, and in all the other bands. The highest power suppression is achieved by the WPTEMD technique in all the bands except the $\gamma$, which is slightly modified by the WPTICA.

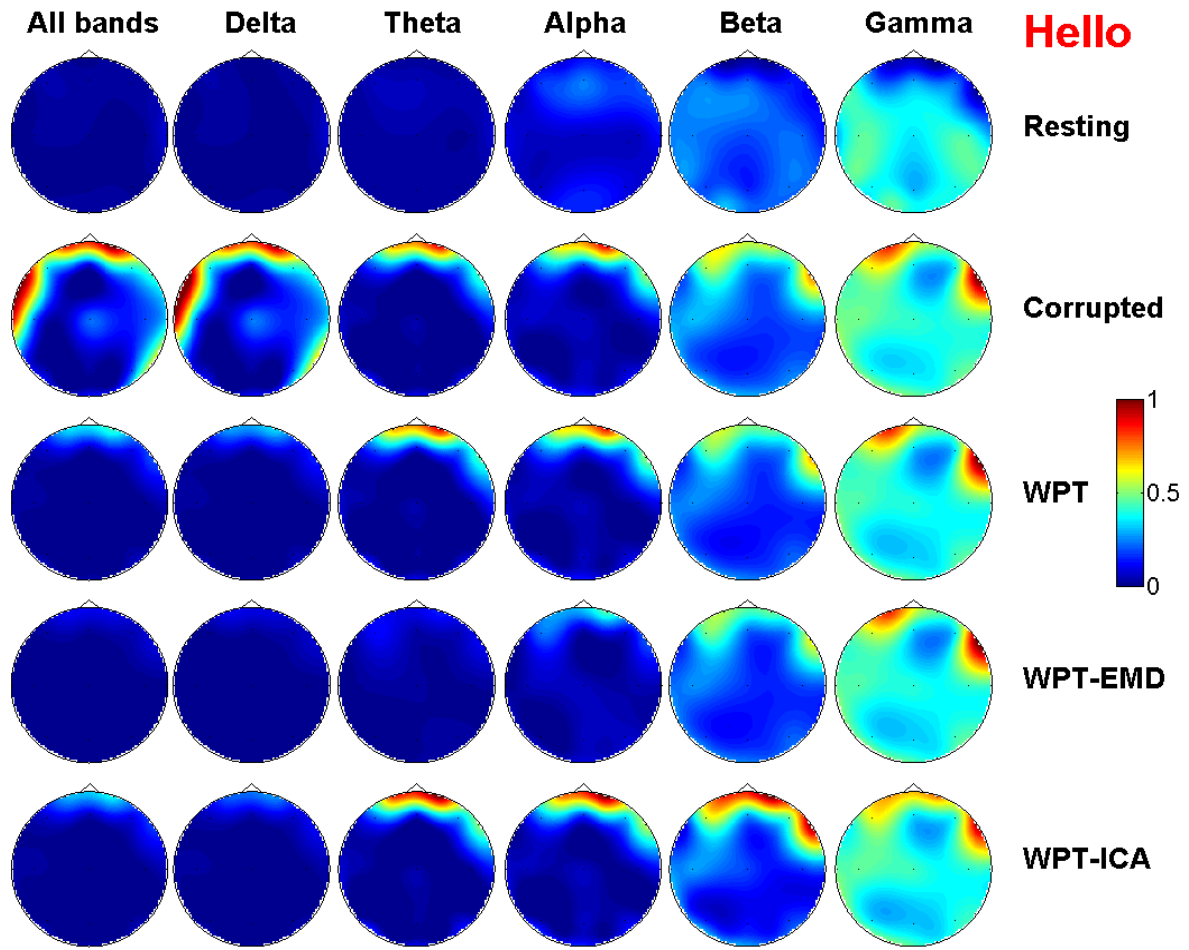

Figure 8: Distribution of the scalp topography power (various bands) for single subject-single trial of saying hello.